\documentclass[prx, tightenlines,10pt,notitlepage,nofootinbib,twocolumn,superscriptaddress]{revtex4-2}

\usepackage{amsmath,epsfig,amssymb}
\usepackage{bbm}
\usepackage{bm}
\usepackage{times}
\usepackage{amsthm}
\usepackage{xcolor}
\usepackage[caption=false]{subfig}
\usepackage{multirow}
\usepackage[normalem]{ulem}
\usepackage{qcircuit}
\usepackage{braket}
\usepackage{amsfonts,amssymb,dsfont}
\usepackage{graphicx}
\usepackage{soul}
\usepackage{hyperref}
\hypersetup{
    colorlinks=true,       
    linkcolor=red,          
  citecolor=magenta,        
    filecolor=magenta,      
    urlcolor=cyan,           
    runcolor=cyan
}

\def\sz{\sigma^z}

\def\Tr{\text{Tr}}

\newcommand{\tr}[1]{{\rm Tr}\left[ #1\right]}

\definecolor{darkgreen}{rgb}{0.0, 0.7, 0.0}

\usepackage{thmtools}
\usepackage{thm-restate}

\begin{document}

\title{Quantifying the Impact of Precision Errors on Quantum Approximate Optimization Algorithms}
\author{Gregory Quiroz}
\email[Corresponding Author:\,]{Gregory.Quiroz@jhuapl.edu}
\affiliation{The Johns Hopkins University Applied Physics Laboratory, Laurel, Maryland, 20723, USA,}

\author{Paraj Titum}
\affiliation{The Johns Hopkins University Applied Physics Laboratory, Laurel, Maryland, 20723, USA}
\author{Phillip Lotshaw}
\affiliation{Quantum Computational Sciences Group, Computational Sciences and Engineering Division, Oak Ridge National Laboratory, Oak Ridge, Tennessee 37831, USA}
\author{Pavel Lougovski}
\thanks{Now at Amazon Web Services}
\affiliation{Quantum Computational Sciences Group, Computational Sciences and Engineering Division, Oak Ridge National Laboratory, Oak Ridge, Tennessee 37831, USA}

\author{Kevin Schultz}
\affiliation{The Johns Hopkins University Applied Physics Laboratory, Laurel, Maryland, 20723, USA}
\author{Eugene Dumitrescu}
\affiliation{Quantum Computational Sciences Group, Computational Sciences and Engineering Division, Oak Ridge National Laboratory, Oak Ridge, Tennessee 37831, USA}
\author{Itay Hen}
\affiliation{Department of Physics and Astronomy, University of Southern California, Los Angeles, California 90089, USA}
\affiliation{Center for Quantum Information Science \& Technology,
University of Southern California, Los Angeles, California 90089, USA}
\affiliation{
Information Sciences Institute, University of Southern California, Marina del Rey, California 90292, USA}

\begin{abstract}

The quantum approximate optimization algorithm (QAOA) is a hybrid quantum-classical algorithm that seeks to achieve approximate solutions to optimization problems by iteratively alternating between intervals of controlled quantum evolution. Here, we examine the effect of analog precision errors on QAOA performance both from the perspective of algorithmic training and canonical state- and observable-dependent QAOA-relevant metrics. Leveraging cumulant expansions, we recast the faulty QAOA as a control problem in which precision errors are expressed as multiplicative control noise and derive bounds on the performance of QAOA. We show using both analytical techniques and numerical simulations that errors in the analog implementation of QAOA circuits hinder its performance as an optimization algorithm. In particular, we find that any fixed precision implementation of QAOA will be subject to an exponential degradation in performance dependent upon the number of optimal QAOA layers and magnitude of the precision error. Despite this significant reduction, we show that it is possible to mitigate precision errors in QAOA via digitization of the variational parameters, therefore at the cost of increasing circuit depth. We illustrate our results via numerical simulations and analytic and empirical error bounds as a comparison. While focused on precision errors, our approach naturally lends itself to more general noise scenarios and the calculation of error bounds on QAOA performance and broader classes of variational quantum algorithms.
\end{abstract}
\maketitle

{\footnotetext[3]{This manuscript has been authored in part by UT-Battelle, LLC, under Contract No. DE-AC0500OR22725 with the U.S. Department of Energy. The United States Government retains and the publisher, by accepting the article for publication, acknowledges that the United States Government retains a non-exclusive, paid-up, irrevocable, world-wide license to publish or reproduce the published form of this manuscript, or allow others to do so, for the United States Government purposes. The Department of Energy will provide public access to these results of federally sponsored research in accordance with the DOE Public Access Plan.}}

\section{Introduction}

The Quantum Approximate Optimization Algorithm (QAOA)~\cite{farhi2014:qaoa} provides a hybrid classical-quantum approach to solving combinatorial optimization problems. QAOA has attracted a great deal of attention due to the advent of near- and intermediate-term quantum computing. Furthermore, because of its simplicity, QAOA has been amenable to both analytical study~\cite{farhi2014:qaoa,farhi2015:qaoa,wecker2016:to,biswas2017:qaoa,jiang2017:gs, wang2018:mc,lloyd2018:qaoa,farhi2019:qs,hadfield2019:qaoa,farhi2020:qaoa} and experimental implementation on a variety of platforms~\cite{kandala2017:vqe,pagano2020:qaoa,bengtsson2020:qaoa,harrigan2021:qaoa}.

While the idealized QAOA provides performance guarantees \cite{farhi2014:qaoa,jiang2017:gs,farhi2020:qaoa}, questions remain regarding its practical implementation, namely its robustness to noise. Previous studies have examined QAOA subject to local decoherence~\cite{alam2019:nqaoa,marshall2020:nqaoa,xue2021:nqaoa} and readout error~\cite{maciejewski2021:qaoa-rd} thus providing some insight into the ramifications of systematic and environment-induced noise on QAOA performance. These studies primarily leveraged numerical investigations to capture dependence on noise parameters. Analytical bounds for QAOA performance and training have been developed for a class of noise models that give a simple rescaling to Pauli terms in the density operator \cite{wang2020:nibp}, e.g., in depolarizing channels. However, to the best of our knowledge, analytical bounds on QAOA success probability or training error in the presence of control noise have yet to be assessed. In this work, we develop bounds particularly focusing on errors in the control parameters. The techniques we develop are amenable to noise generated by imperfect control Hamiltonians, yet extendable to more generic noise models. Without loss of generality we illustrate their use within a simple yet representative error model. 

QAOA is well-defined in the gate-model setting, however, its practical implementation involves the alternating application of unitary operators that depend on continuous parameters that must be optimized. Cases exist where the number $p$ of such alternations, commonly referred to as the \emph{order} of the algorithm, can remain small while achieving an approximately satisfactory solution~\cite{farhi2014:qaoa}. Unfortunately, a majority of such cases do not coincidence with problem classes of practical importance. As such, typically the QAOA order must increase commensurately with the problem size (and clause density when applicable) to generate near-optimal configurations~\cite{guerreschi2019:QAOA,niu2019:QAOA,akshay2020:rd,akshay2020:rd2,willsch2020:QAOA}. 



In practical settings, the increasingly large sets of continuous variables upon which QAOA is founded are prone to misspecification in which the implemented parameters of the QAOA differ from those intended. We designate these errors as \emph{precision errors} in the variational parameters and model these as imperfect control in the QAOA Hamiltonians. While static precision errors can be absorbed as a constant shift in the optimized parameters, temporally drifting errors varying on timescales shorter than the total algorithm runtime pose a more significant threat to the success of the algorithm. This is most notably true if all parameters cannot be retrained more rapidly than the characteristic timescale of the error. Time-varying precision errors with short correlation times can quickly accumulate in optimization problems requiring large $p$, ultimately spoiling the computation and resulting in errors in parameter training and unfavorable reductions in performance guarantees.

The deleterious consequences of precision errors were recognized early on in gate-model quantum computing~\cite{landauer1995:quant} and more recently in Hamiltonian quantum computing, in particular quantum annealing (QA)~\cite{albash2019:ae,pearson2019:doom}. Although the former may be addressed theoretically via quantum error correction~\cite{shor1996:ftqc},
questions remain regarding how detrimental such errors are to NISQ-era algorithms and how to effectively combat them without error correction. Moreover, while QA has shown improvement in performance through quantum annealing correction, the presence of unmitigated precision errors has proven to be severely harmful~\cite{pearson2019:doom}. Misspecification of programmable, continuous parameters within the cost function leads to exponential decay in success probability with problem size and error magnitude~\cite{albash2019:ae}. To date, it is unclear whether this behavior observed for QA is indeed a bad omen for QAOA, given the intimate relationship between the two paradigms ~\cite{farhi2014:qaoa}.



\begin{figure*}[t]
    \centering
    \includegraphics[width=\textwidth]{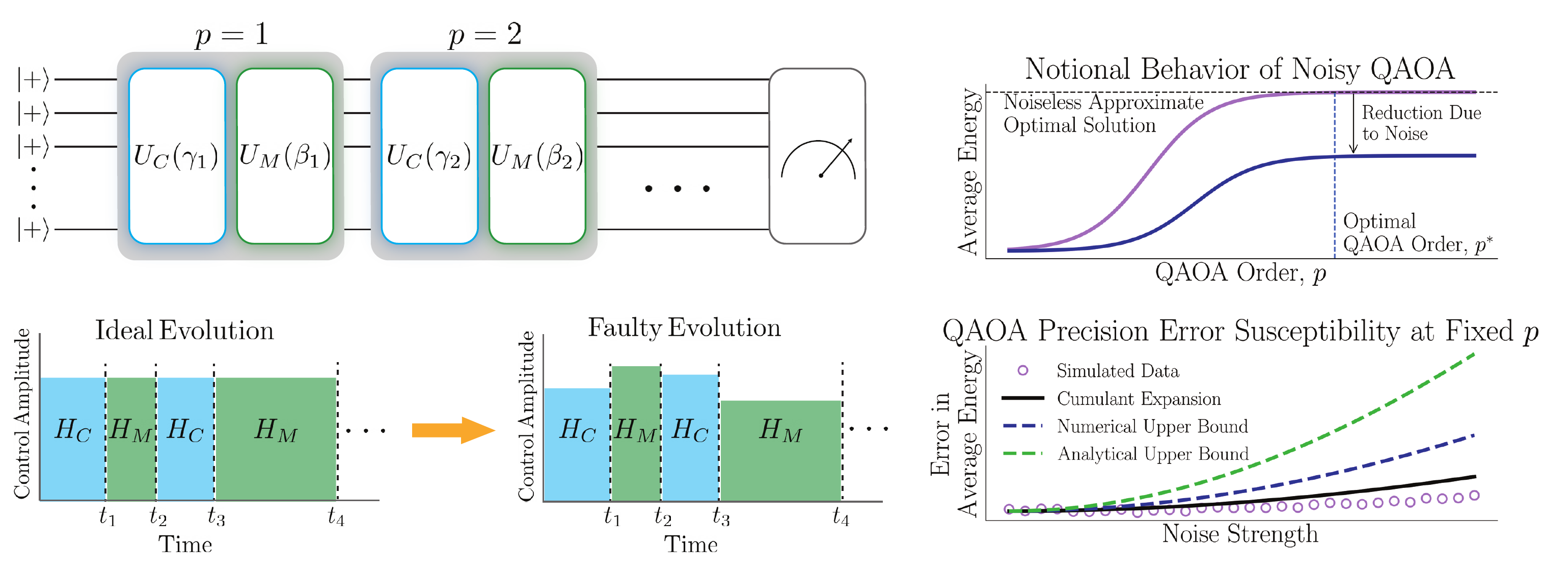}
    \caption{Schematic figure describing the effect of faulty QAOA evolution resulting from precision errors. Top-left: Schematic of typical QAOA ans\"atze. Bottom-left: Perspective of precision errors on QAOA evolution. Ideally, each Hamiltonian is implemented with a specific amplitude and duration. Precision errors in the variational parameters can be modeled as amplitude errors in the QAOA control evolution. Top-right: Notional behavior of QAOA in the ideal and noisy setting. As the QAOA order is increased, the algorithm performance improves. Ultimately, performance plateaus at the approximate optimal solution, where an optimal QAOA order can be determined. The presence of noise alters this behavior, leading to a reduction in the optimal approximate solution. Bottom-right: Utility of cumulant-based perspective on faulty QAOA illustrated for notional QAOA performance in the presence of precision errors. The cumulant expansion provides an analytical expression for the approximate dynamics of the faulty QAOA, as well as error bounds on QAOA performance.}
    \label{fig:overview}
\end{figure*}

In this study, we address these concerns via an analytical and numerical investigation of QAOA in the presence of precision errors. Providing insight into the extent of their harm on parameter training and performance guarantees, we enable the development of more robust estimates of QAOA complexity. Our approach relies on concepts from quantum control theory to analytically estimate the contribution of precision errors. This approach paves the way for the development of bounds on approximation ratios and training error that focus on precision errors and also enable further insight into broader classes of noise models. Through our analysis, we argue that any fixed precision implementation of QAOA is destined to detrimentally affect the success of QAOA, which represents a fundamental limitation to the scalability of the algorithm, and more broadly, Ising machines. In particular, we find an \emph{exponential reduction} in success probability with increasing problem size and error magnitude. This observation is displayed through numerical investigations of the QAOA variant of Grover's search and the one-dimensional transverse-field Ising model. Despite the apparent ``doomsday scenario" for QAOA, we show that it is possible to mitigate precision errors in QAOA via digitization of the variational parameters.

The manuscript is organized as follows. In Sec.~\ref{sec:summary} we provide a concise summary of the main results, and bounds on the performance of QAOA for optimization. In Sec.~\ref{sec:QAOA} we provide a general overview of the QAOA algorithm and the model of precision errors considered in this paper. In Sec.~\ref{sec:error-dyn} we derive the general approach based on the cumulant expansion to analyze the error in QAOA evolution due to precision errors. In Sec~\ref{sec:err-an}, we derive expressions for the cumulant for the specific case of coherent and stochastic errors before deriving bounds on the errors in terms of difference in expectation values, approximation ratio, gradients and unitaries in Sec.~\ref{sec:bounds}. In Sec.~\ref{sec:examples} we show numerical evidence of the effects of precision errors in two example QAOA algorithms implementing (i) Grover search and (ii) 2-SAT (Ising) on a ring. Finally in Sec.~\ref{sec:digitization} we discuss an approach to mitigating the effects of precision errors in the implementation of QAOA.






\section{Summary  of Main Results}
\label{sec:summary}
Variational training parameters in QAOA are susceptible to precision errors that potentially threaten the success of the algorithm. We model precision errors via a control noise contribution to the QAOA Hamiltonians. This results in a misspecification of the QAOA control amplitudes, as depicted in the bottom-left panel of Fig.~\ref{fig:overview}. Mathematically, parameter misspecifications are captured by time-dependent error functions that are specified by white noise processes or constant errors to model precision errors in two specific regimes. 
Upon subjecting QAOA to precision errors, we observed an exponential decrease in the achievable approximate optimal solution for QAOA with increasing noise strength and optimal QAOA order; see Fig.~\ref{fig:overview}(bottom-right) for an illustration of this effect. 

Using concepts from quantum control theory, we quantify the effect of precision errors on expectation values and unitary operators. Recent work in quantum control has sought to utilize cumulant expansions to assess interplay between control schemes and temporally-correlated noise processes~\cite{pazsilva2014:fff, pazsilva2017:fff, norris2018:cn}. Inspired by this work, we leverage a cumulant expansion to examine the dynamics of a system subject to a control designated by the noiseless QAOA protocol and noise produced by a control error Hamiltonian representing precision errors. The cumulant-based approach affords a novel perspective on faulty QAOA dynamics that facilitates the development of bounds on various QAOA-relevant quantities. In particular, our approach allows for the development of bounds on approximation ratios and parameter training error. Fig.~\ref{fig:overview}(bottom-right) illustrates the utility of our approach for assessing QAOA susceptibility to precision errors. The cumulant expansion yields expressions for the approximate dynamics of the noisy QAOA. Upper bounds on QAOA-relevant quantities (e.g., the error in average energy) are determined via the cumulant approach. Together, these expressions capture the relative behavior of QAOA as a function of noise strength and QAOA order.

To benchmark our analysis we numerically study the performance of QAOA in the presence of precision errors in two specific implementations: 
\begin{itemize}
    \item [(i)] {\it Grover Search}: We implement the analytically obtained optimal Grover QAOA for identifying a marked state $\ket{0}$ from a database of $N=2^n$ entries~\cite{Jiang2017}. The optimal circuit has $p^*\sim \sqrt{N}$ layers.
    \item [(ii)] {\it Ising instances}: We utilize optimal circuits to prepare $n$ qubit GHZ states using QAOA with a Ising Hamiltonian on a ring geometry~\cite{Ho_2019}. The optimal circuit has $p^*=n/2$ layers. 
\end{itemize}
In both cases, $p^*$ is the optimal number of QAOA layers and $n$ indicates the number of qubits. We numerically simulate the QAOA circuits for these two problem instances and calculate the measured errors in the cost observable as well as the average distance between the noisy and perfect unitary operators. The precision errors are drawn from a normal distribution, $\mathcal{N}(\eta,\sqrt{\Gamma})$ with mean $\eta$ and variance $\Gamma$. Specifically, we consider two representative limiting cases: (a) Stochastic errors: $\eta=0$, drawn from $\mathcal{N}(0,\sqrt{\Gamma})$, and (b) Coherent errors: $\Gamma=0$, drawn from $\mathcal{N}(\eta,0)$. 

Finally, let us summarize the main numerical results as well as the analytical bounds obtained on the faulty QAOA evolution as a result of precision errors. Let us consider a QAOA circuit with a total runtime $T$ (and layers $p$) and perfect implementation leading to the state $\rho_0(T)$.
\begin{enumerate}
    \item {\it Expectation values}-- The noise-averaged error in the measured expectation value of any observable, $O$ has the following upper bound, 
    \begin{equation}
    |\overline{\Delta\braket{O(T)}}| \leq \left(e^{\|\mathcal{C}_O(T)\|_\infty}-1\right) \|O \rho_{0}(T)\|_1,
    \label{eq:sor-exp-bound}
    \end{equation}
    where $\mathcal{C}_O(T)$ [defined in Eq.~(\ref{eq:COdef-cumulant-expansion})] denotes the cumulant expansion for the error operator $\Lambda(T)$ that depends on the noise as well as the observable, $O$; see Eq.~(\ref{eq:avg_O_lamb}). The operator $\mathcal{C}_O(T)$ is dependent upon the specifications of the problem; however, it can be shown to scale as $\|\mathcal{C}_O(T)\|_\infty \lesssim \mathcal{O}(\eta p^*)$ and $\|\mathcal{C}_O(T)\|_\infty \lesssim \mathcal{O}(\Gamma p^*)$ for constant and stochastic precision errors, respectively. The numerical investigations of QAOA circuits for both examples considered, indicate the following scaling for the error in expectation value of the cost function Hamiltonian, $H_C$ for weak noise,
    \begin{align}
        |\Delta H_C|&\approx \braket{H_C}_0 \left(1-\exp\left[-\chi_{H}(p^*)\right]\right),\\
        \textrm{with, } \chi_{H}(p^*) &\equiv\begin{cases}
        \chi(\Gamma p^*), & \textrm{stochastic error}\\
        \chi(\eta^a p^*), & \textrm{coherent error}
        \end{cases}
    \end{align}
    where $\braket{H_C}_0$ is the noiseless expectation value and the exact functional form of $\chi_H(p^*)$ depends on the details of the QAOA algorithm. The effects of the coherent error turns out to be dependent on the algorithm, with the $a=1$ for Grover search and $a=2$ for the Ising problem.
    \item {\it Unitary operators}-- We obtain a bound on the operator norm of the difference between the perfect and faulty unitary operators for the QAOA evolution via the Frobenius norm,
    \begin{align}
        \overline{\|U-U_0\|}^2_\infty \leq \overline{(\Delta U)^2} \leq 2\left(e^{\|\mathcal{C}(T)\|_\infty}-1\right),
    \end{align}
    where $\Delta U=\|U-U_0\|_2$ and $\mathcal{C}$ denotes the cumulant expansion for the error operator. [Note that the difference between unitaries has a direct relationship with the error in the expectation value; see Eq.~(\ref{eq:app-bias-bound-const}).] Numerically, we evaluate the $\infty$-norm difference between the QAOA evolution operators for the Grover search algorithm. For weak noise and small number of layers $p$ the average difference scales as,
    \begin{align}
            \overline{\|U-U_0\|}_\infty &\propto
    \begin{cases}
      \sqrt{\Gamma p},& \textrm{[stochastic error]} \\ 
     \eta p, &\textrm{[coherent error]} \\ 
    \end{cases}
    \end{align}
    For large number of layers, or equivalently large noise, the difference saturates to a maximum value $2$. 
    \item {\it Training Error}-- We derive bounds on the error in training variational parameters for gradient-based and closed-loop optimization routines. The former follows from the absolute error in the gradient of the expectation value, which can be bounded by expressions proportional to the operator norm (largest singular value) of the error operator $\Lambda(T)$ and its gradient. We estimate training error in closed-loop optimization via bounds on the mean-squared error (MSE) in the expectation value of $H_C$. In the case of constant errors, the training error is bounded by
    \begin{equation}
        {\rm MSE}(H_C(T))\leq 2C^2_{\rm max} \left(1 + 2 h^2_E\right),
    \end{equation}
    where $C_{\rm max}$ is the operator norm of $H_C$ (or equivalently, the maximum value of the cost function) and $h_E$ effectively represents the accumulated precision error; see Eq.~(\ref{eq:mse-coherent}). Similar expressions that depend more generically on the operator norm of $\Lambda(T)$ are obtained for stochastic precision errors.
    \item {\it Approximation ratio}-- The bound on the noise-averaged error in the expectation value naturally leads to the bound on the error in the optimal approximation ratio
    \begin{align}
        |\overline{\Delta \epsilon^*}| \leq \frac{1}{C_{\rm max}} \left(e^{\|\mathcal{C}_O(T^*)\|_\infty}-1\right) \|H_C\, \rho_{0}(T^*)\|_1.
    \end{align}
    The algorithmic runtime $T^{*}$ denotes the time required to achieve the noiseless optimal approximation ratio $\epsilon^{*}$. The Hamiltonian $H_C$ denotes the cost function Hamiltonian.
\end{enumerate}

Through our analytical and numerical analysis, we find that stochastic precision errors yield exponentially increasing error in expectation values and distance between unitaries. This observation subsequently propagates to parameter training and approximation ratios. Despite the detrimental nature of precision errors, we devise a strategy for effectively addressing them by digitizing the variational parameters in a binary representation and implementing each QAOA evolution operator as a composite operator. This strategy relies on each constituent operator possessing a greater precision than the desired precision of the QAOA. More precisely, for a $p$-layer QAOA to achieve a desired accuracy of $2^{-\epsilon}$, the precision error can be at most $\sim \epsilon/\sqrt{p}$. Provided that such stipulations can be met, the digitization strategy ultimately requires an increase in the circuit depth by a factor $\propto |\log(\epsilon/\sqrt{p})|$. Digitization is not required for constant coherent errors, where the over- or under-rotation can be accounted for (1) by updating the optimized angles exactly by the coherent error or (2) intrinsically by a closed-loop classical optimization routine used for parameter training.  

\section{Quantum Approximate Optimization Algorithm}
\label{sec:QAOA}
In combinatorial optimization, approximation algorithms seek solutions with provable guarantees on the distance between the value of the returned solution and the global optimum. More concretely, consider an objective function $C:\{0,1\}^n\rightarrow \mathbb{R}$ which is to be optimized. An $\epsilon^*$-approximation algorithm with high probability achieves a solution $x^*$ such that
\begin{equation}
    \frac{C(x^*)}{C_{\rm max}}\geq \epsilon^*.
\end{equation}
Under these conditions, the algorithm insures with high probability that all solutions are within the approximation ratio $\epsilon^*$ of the maximum $C_{\rm max}$.

The QAOA generates approximate solutions by stroboscopically alternating between two intervals of controlled evolution. In QAOA, these evolutions are generated by a mixing Hamiltonian $H_M$ and a problem Hamiltonian $H_C$. The mixing Hamiltonian can take various forms; however, typically, it assumes the form of a global transverse field
\begin{equation}\label{mixing H}
    H_M = \sum^n_{j=1} X_j.
\end{equation}
Other mixing Hamiltonians that account for problem constraints have also been examined~\cite{hadfield2019:qaoa}.

The problem Hamiltonian is constructed by translating the classical cost function of $n$-binary variables into a Hamiltonian of $n$ qubits. This is accomplished by introducing the variable transformation $x_i=(1-z_i)/2$, where $z_i\in\{-1,1\}$. The cost function $C(z)$ is then transformed into $H_C$ by replacing $z_i$ with the Pauli operator $\sz_i$. The resulting problem Hamiltonian is an Ising Hamiltonian defined on the underlying graph of the cost function $C(z)$. Note that $H_C$ acts diagonally on computational basis states $\ket{z}$, i.e.
\begin{equation}
    H_C\ket{z}=C(z)\ket{z}.
\end{equation}

The QAOA is formally implemented via a time-dependent Hamiltonian
\begin{equation}
    H_{0}(t) = [1-s(t)] H_M + s(t) H_C,
    \label{eq:H-QAOA}
\end{equation}
where 
\begin{equation}
    s(t)=\left\{
    \begin{array}{lcr}
         1 & : & t\in \Delta t^C_j\\
         0 & : & t\in \Delta t^M_j
    \end{array}
    \right.,
\end{equation}
defines a stroboscopic function that toggles between weighted time-evolutions generated by $H_M$ and $H_C$. The time intervals $\Delta t^{C}_j=[t_{2j-2},t_{2j-1}]$ and $\Delta t^{M}_j=[t_{2j-1},t_{2j}]$, $j=1,\ldots, p$ define the periods during with $H_C$ and $H_M$, respectively, govern the evolution. The dynamics generated by $H_{0}$ can be expressed as
\begin{eqnarray}
    U_{0}(T,0) &=& \mathcal{T}_+ e^{-i\int^{T}_0 H_{0}(t) dt}\nonumber \\
    &=&U_M(\beta_p) U_{C}(\gamma_p) \,\cdots\, U_M(\beta_1)U_{C}(\gamma_1), \nonumber \\
    &=& U^{(p)}_0(\bm{\gamma}, \bm{\beta}),
    \label{eq:QAOA}
\end{eqnarray}
where $\mathcal{T}_+$ denotes the time-ordering operator. The variational parameters $\bm{\gamma}=(\gamma_1,\ldots,\gamma_p)$ and $\bm{\beta}=(\beta_1,\ldots,\beta_p)$ control the time-intervals over which $H_M$ and $H_C$ are applied. Each $(\gamma_j,\beta_j)$ are real numbers that parameterize the evolution operators $U_M(\beta)=e^{-i\beta H_M}$ and $U_C(\gamma)=e^{-i\gamma H_C}$ and characterize the total algorithmic runtime $T=\sum^{p}_{j=1} (|\gamma_j| + |\beta_j|)$.

QAOA evolution is applied to a state initialized in an equal superposition state between all computational basis states, or equivalently the ground state of $-H_M$. The initial state is explicitly given by
\begin{equation}
    \rho(0) = \ket{\psi(0)}\bra{\psi(0)} = \ket{+}\bra{+}^{\otimes n},
\end{equation}
where $\ket{+}=1/\sqrt{2}(\ket{0}+\ket{1})$ is defined by the single qubit computational basis states $\{\ket{0},\ket{1}\}$. The state resulting from the QAOA evolution is then given by
\begin{equation}
    \rho_{0}(T) = U^{(p)}_{0}(\bm{\gamma}, \bm{\beta})\, \rho(0)\, U^{(p)}_{0}(\bm{\gamma}, \bm{\beta})^\dagger.
\end{equation}

The variational parameters are optimized using a classical optimization routine. While there are many options for this procedure~\cite{qiskit-textbook, zhueaaw2019:qaoa, streif2020:qaoa, wauters2020:rl}, all of them seek to obtain an approximate solution to the optimization problem by maximizing
\begin{eqnarray}
    F(\bm{\gamma}, \bm{\beta}) &=& \braket{H_C}_{\bm{\gamma}, \bm{\beta}} \nonumber\\
    &=& \Tr\left[\rho_{0}(T) H_C\right],
    \label{eq:obj}
\end{eqnarray}
the average energy with respect to the problem Hamiltonian. One can define the approximation ratio for the QAOA with respect to Eq.~(\ref{eq:obj}) as
\begin{equation}
    \epsilon \equiv F(\bm{\gamma}, \bm{\beta})/C_{\rm max}.
\end{equation}
The QAOA is now said to be an $\epsilon^{*}$-approximation algorithm if $\epsilon\geq \epsilon^{*}$. The approximation condition is guaranteed if the resulting quantum state $\rho_{0}(T)$, when measured in the computational basis, is highly concentrated on solutions that are $\epsilon^{*}$-approximately optimal~\cite{farhi2014:qaoa}. 

\subsection{Precision Errors}
Control Hamiltonians are designed to implement a particular quantum operation. However, physical realizations of these Hamiltonians are often accompanied by undesired control-dependent additive or multiplicative noise~\cite{ball2015:walsh,frey2017:cn,norris2018:cn}. As a result, unwanted system dynamics generated by control parameter inaccuracies lead to control errors. Precision errors constitute a particular type of control error that while being associated with the misspecifications of variational parameters, also can be treated as control amplitude errors in the QAOA evolution. It is this quantum control point of view that we utilize to examine precision errors in the QAOA setting.



Precision errors can be modeled in QAOA by unintended deviations in the variational parameters. This is equivalent to introducing control noise into the QAOA evolution, such that the faulty dynamics are governed by
\begin{equation}
    H(t) = H_{0}(t) + H_E(t).
    \label{eq:H-f-QAOA}
\end{equation}
The dynamics are now governed by the ideal QAOA and an additional error Hamiltonian
\begin{equation}
    H_E(t) = \eta_M(t)[1-s(t)] H_M + \eta_{C}(t) s(t) H_{C}.
    \label{eq:errors}
\end{equation}
The error functions $\eta_M(t)$ and $\eta_{C}(t)$ capture the time-dependent control errors for the mixer and problem Hamiltonians, respectively. The evolution generated by Eq.~(\ref{eq:H-f-QAOA}) is that of a faulty QAOA evolution $U^{(p)}_0(\bm{\tilde{\gamma}}, \bm{\tilde{\beta}})$, where  $\tilde{\gamma_j}=\gamma_j(1 + \eta_{C,j})$ and $\tilde{\beta_j}=\beta_j(1 + \eta_{M,j})$ are implemented rather than the ideal evolution of Eq.~(\ref{eq:QAOA}). So far our model 
describes a general class of 
multiplicative control errors, i.e., errors that are generated by a faulty control Hamiltonian where the error magnitude scales proportionally with the control.

The statistical properties of the error functions  further specify the precision error characteristics. Namely, we model the error functions as stationary, Gaussian random variables with mean, $\overline{\eta_\mu(t)}=\eta_\mu$, and time-dependent two-point correlation functions 
\begin{eqnarray}
    \overline{\eta_\mu(t_1) \eta_\nu(t_2)} &=& \delta_{\mu\nu}\Gamma_\mu(t_1-t_2).
    \label{eq:noise-props}
\end{eqnarray}
where $\Gamma_\mu(t_1-t_2)$ describes the autocorrelation function for $\mu,\nu \in \{M,C\}$. Note that $\overline{\cdots}$ denotes classical ensemble averaging. Below, we will focus on two cases: (1) constant coherent errors, where $\Gamma_\mu\equiv \eta^2_\mu$ and (2) stochastic errors, where $\Gamma_\mu(t_1-t_2) = \Gamma_\mu \delta(t_1-t_2)$. Together, these scenarios capture two extremes of precision errors, each uniquely detrimental to QAOA with distinct scaling behavior in QAOA performance as a function of $p$. 

\section{Error Dynamics}
\label{sec:error-dyn}
We investigate the effect of precision errors on QAOA performance via the average energy with respect to the problem Hamiltonian. The faulty QAOA dynamics generated by Eq.~(\ref{eq:H-f-QAOA}) leads to errors in the final state of the system. Ultimately, these errors propagate to the objective function given in Eq.~(\ref{eq:obj}). In order to understand the consequences of precision errors on $F(\bm{\gamma},\bm{\beta})$, we leverage concepts from control theory to isolate the error dynamics dictated by $H_E(t)$. We examine the error dynamics through a cumulant expansion approach that has been previously employed for general studies of open quantum system dynamics in the presence of temporally correlated noise~\cite{pazsilva2017:fff}.

\subsection{Rotated-Frame Dynamics}
In quantum control, time-dependent perturbation theory (TDPT) has proven to be an invaluable tool for examining weak noise processes in the presence of strong control. TDPT expansions are typically performed by moving into a rotating reference frame with respect to the control -- commonly referred to as the \emph{toggling-frame}. The noise contributions are then studied within this rotating frame using a particular expansion technique, e.g., Dyson or Magnus expansion~\cite{viola1999:dd,khodjasteh2005:dd}.

Utilizing the framework of quantum control, the faulty QAOA is analyzed through the lens of TDPT. We work under the assumption that the dynamics are dominated by the ideal QAOA evolution and thus impose a strong control condition $\|H_E(t)\|\ll \|H_{0}(t)\|$ $\forall t$. Furthermore, we enforce a weak noise assumption $\|H_E(t)\|\ll 1$. Together, these conditions enable the effects of $H_E(t)$ on the QAOA evolution to be analyzed via TDPT. 

The first step towards isolating the effects of $H_E(t)$ is to move into an interaction picture co-rotating with the ideal evolution $H_{0}(t)$. The erroneous Hamiltonian transforms into 
\begin{eqnarray}
    \tilde{H}_E(t) &=& U_{0}(T,t) H_E(t) U^\dagger_{0}(T,t)\nonumber \\
    &=& \eta_M(t)[1-s(t)] \tilde{H}_M(t) + \eta_{C}(t) s(t) \tilde{H}_{C}(t). \quad\quad
\end{eqnarray}
The mixer and problem Hamiltonian in the rotated frame are given by $\tilde{H}_M(t)=U_{0}(T,t)H_M U^\dagger_{0}(T,t)$ and $\tilde{H}_C(t)=U_{0}(T,t)H_C U^\dagger_{0}(T,t)$, respectively. Note that the rotated error Hamiltonian is expressed in terms of a ``reverse" interaction picture with respect to the ideal QAOA evolution. As a result, the total dynamics evolution operator
\begin{equation}
    U(T) = \mathcal{T}_{+}e^{-i\int^{T}_0 H(t)dt}
    \label{eq:u-tot}
\end{equation}
is factorized into the product
\begin{equation}
    U(T) = \tilde{U}_E(T) U_{0}(T),
    \label{eq:u-rep}
\end{equation}
where
\begin{equation}
    \tilde{U}_{E}(T) = \mathcal{T}_{+}e^{-i\int^{T}_0 \tilde{H}_{E}(t)dt}.
    \label{eq:u-err}
\end{equation}
and $U_0(t)$ is the ideal evolution generated by Eq.~(\ref{eq:H-QAOA}). The operator $\mathcal{T}_{+}$ denotes time-ordering. The advantage of this formulation will become clear in the next section and the equivalence between this approach and the canonical rotated frame is discussed in Appendix~\ref{sec:alt-toggling}.

\subsection{Time-dependent Dynamics of Observables}
Expectation values are commonly utilized in variational quantum algorithms to evaluate algorithmic performance and carry out variational parameter training. In QAOA, the average energy with respect to the problem Hamiltonian is the expectation value of interest as in Eq.~(\ref{eq:obj}). This metric is utilized to train variational parameters, to maximize the objective function defined by $H_C$, which in turn defines the approximation ratio for a particular problem and depth. Therefore, the effect of noise on this expectation value will ultimately influence QAOA parameter training and bounds on performance guarantees.

In order to investigate the effect of precision errors on QAOA performance, we first consider the case of a general observable $O$. The noise-averaged, time-dependent dynamics of $O$ subject to the faulty QAOA are described by 
\begin{equation}
    \overline{\braket{O(T)}}= \overline{\braket{\text{Tr}\left[\rho(T)O\right]}}
    \label{eq:avg-o}
\end{equation}
where it is assumed that we are particularly interested in the expectation value of $O$ after the total algorithmic runtime $T$. 
Using 
\begin{equation}
    \rho(t) = \tilde{U}_E(t) U_{0}(t)\rho(0) U^\dagger_{0}(t)\tilde{U}^\dagger_E(t),
\end{equation} the expectation value can be rewritten as
\begin{eqnarray}
    \overline{\braket{O(T)}}
    &=&\overline{\braket{\text{Tr}\left[\tilde{U}_E(T) \rho_{0}(T)\tilde{U}^\dagger_E(T) O\right]}},\nonumber\\
    &=&\text{Tr}\left[\Lambda(T) \rho_{0}(T) O\right].
    \label{eq:avg_O_lamb}
\end{eqnarray}
The reason for the particular rotating frame choice is now evident: it allows for the dynamics to be partitioned into an error operator $\Lambda(T)=\overline{O^{-1}\tilde{U}^\dagger_E(T)O\tilde{U}_E(T)}$ and the ideal time-evolved state $\rho_{0}(T)$. The error operator encapsulates all of the error dynamics, i.e., all dynamics generated by the error Hamiltonian. Eq.~(\ref{eq:avg_O_lamb}) expresses the error dynamics in an intuitive way, where an absence of $H_E(t)$ results in $\Lambda(T)=I$. This perspective can be quite useful, especially when one seeks to mitigate the effects of the noise, and thus, minimize the distance between $\Lambda(T)$ and the identity operator. It is important to note that $\Lambda(T)$ demands $O$ to be invertible, however, as we discuss in Appendix~\ref{subsec:noninv-o}, if one can write a general observable $O$ as a sum of invertible operators then similar expression to Eq.~(\ref{eq:avg_O_lamb}) can be obtain. 

Formally, the error operator can be expressed in terms of a cumulant expansion. This is observed by rewriting the error operator as
\begin{eqnarray}
    \Lambda(T) 
    &=&\overline{\mathcal{T}_+e^{-i\int^T_{-T} \tilde{H}_{E,O}(t)dt}},
    \label{eq:lambda-cum}
\end{eqnarray}
where the observable-dependent effective Hamiltonian is given by
\begin{equation}
    \tilde{H}_{E,O}(t)=\left\{
\begin{array}{cl}
    \tilde{H}_E(T+t) & t\in[-T,0],\\
    -O^{-1} \tilde{H}_E(T-t)O & t\in[0,T]
\end{array}\right..
\end{equation}
Eq.~(\ref{eq:lambda-cum}) lends itself to the cumulant expansion in which the cumulant expressions can be obtained from the moment-generating equation
\begin{equation}
    \Lambda(T)=
    \overline{\mathcal{T}_+e^{-i\int^T_{-T} \tilde{H}_{E,O}(t)dt }} =
    e^{\mathcal{C}_O(T)}, \label{eq:COdef-cumulant-expansion}
\end{equation}
where $\mathcal{C}_O(T)=\sum^\infty_{k=1}(-i)^k\frac{\mathcal{C}^{(k)}_O(T)}{k!}$. In general, the expansion of $\Lambda(T)$ involves an infinite number of terms. While in certain scenarios the above cumulant expansion can be truncated \emph{exactly}~\cite{pazsilva2017:fff}, this will not be possible for the faulty QAOA problem. As a result, we will truncate the expansion to capture the \emph{approximate} error dynamics. Leveraging the weak noise condition discussed above, the error dynamics are assumed to be well-approximated by a second order truncation on $\Lambda(T)$. More concretely, it is assumed that
\begin{eqnarray}
    \left\|\mathcal{C}_O(T)\right\|_\infty \simeq \|\mathcal{C}^{(1)}_O(T) +i\frac{1}{2}\mathcal{C}^{(2)}_O(T)\|_\infty,
\end{eqnarray}
where $\|\cdot\|_\infty$ is the operator norm, or largest singular value. Note that through the triangle inequality, it is trivial to show that the truncated norm of the cumulant expansion can be upper bounded as $\|\mathcal{C}_O(T)\|_\infty\lesssim \|H_E(T)\|_\infty + \|H_E(T)\|^2_\infty$; thus, explicitly conveying a connection between the cumulant expansion and the weak noise condition.

Under the above condition, the error dynamics are characterized by the first and second cumulants. The first order term 
\begin{equation}
    \mathcal{C}^{(1)}_O = \int^{T}_0 dt_1 \left[C^{(1)}(\tilde{H}_E(t_1)) - C^{(1)}(\tilde{H}^\prime_{E}(t_1))\right]
    \label{eq:c1}
\end{equation}
is dependent upon the rotated-frame error Hamiltonian, its observable-conjugated counterpart $\tilde{H}^\prime_{E}(t)=O^{-1} \tilde{H}_{E}(t)O$, and the cumulant expression $C^{(1)}(A) = \overline{A}$. The second order term

\begin{equation}
    \frac{\mathcal{C}^{(2)}_O}{2!} = I_1(T) + I_2(T) - I_3(T) - I_4(T),
    \label{eq:c2}
\end{equation}
can be decomposed into four integrals:
\begin{eqnarray}
   I_1(T)&=&\int^{T}_0 dt_1\int^{t_1}_0 dt_2\,\,C^{(2)}(\tilde{H}_E(t_1),\tilde{H}_E(t_2))\\
   I_2(T)&=&\int^{T}_0 dt_1\int^{t_1}_0 dt_2\,\,C^{(2)}(\tilde{H}^\prime_{E}(t_2),\tilde{H}^\prime_{E}(t_1))\\
   I_3(T)&=&\int^{T}_0 dt_1\int^{t_1}_0 dt_2\,\,C^{(2)}(\tilde{H}^\prime_{E}(t_1),\tilde{H}_E(t_2))\\
   I_4(T)&=&\int^{T}_0 dt_1\int^{t_1}_0 dt_2\,\,C^{(2)}(\tilde{H}^\prime_{E}(t_2),\tilde{H}_E(t_1)).
\end{eqnarray}
Each integral includes the second order cumulant expression
\begin{equation}
    C^{(2)}(A,B) = \overline{AB} - \frac{1}{2}\left(\overline{A}\,\overline{B} + \overline{B}\,\overline{A}\right).
\end{equation}
Together, these equations form the quantum control framework that will be exploited to calculate the approximate error dynamics generated by $H_E(t)$ and subsequently, the expectation value of $H_C$ in the presence of precision errors. Furthermore, below it will be shown that the cumulant expressions facilitate the development of error bounds on various QAOA-relevant metrics.

\section{Cumulant Analysis}
\label{sec:err-an}
The truncated cumulant expansion can be utilized to calculate the approximate error dynamics of QAOA subject to precision errors. Below, we investigate the dynamics of QAOA for coherent and stochastic errors on the variational parameters.

\subsection{Constant Coherent Errors}
Constant coherent errors are trivially captured by the cumulant expansion, most notably when the error mean is small, i.e., $\eta_\mu\ll 1$. In this case, the dynamics are well-characterized by the first cumulant, with the second order term providing an additional correction that becomes more relevant with increasing error mean. Assuming the QAOA coherent error is weak, we find the error dynamics to be governed by
\begin{eqnarray}
    \mathcal{C}^{(1)}_O(T) &=& \sum_{\mu=M,C}\eta_\mu\int^{T}_0 dt\left(\tilde{H}_\mu(t) - O^{-1} \tilde{H}_\mu(t)O\right)\quad.\nonumber \\
\end{eqnarray}
The QAOA evolution is piece-wise constant which allows for the dynamics to be partitioned into a sum of integrals in which the control propagator can be expressed as
\begin{equation}
    U_0(T,t) = \left\{
    \begin{array}{lcr}
    Q_{p:j+1}e^{-i(t_{2j}-t)H_M} &:& t\in\Delta t^{\text{M}}_j\\
    Q_{p:j}U^\dagger_{C}(\gamma_j)e^{-i(t_{2j-1}-t)H_C} &:& t\in\Delta t^{\text{C}}_j
    \end{array}\right.,
    \label{eq:uc-explicit}
\end{equation}
where 
\begin{equation}
    Q_{k:j} = U_M(\beta_k)U_{C}(\gamma_k) \cdots U_M(\beta_j)U_{C}(\gamma_j).
\end{equation}
Examining the error Hamiltonian and the control propagator, one finds that certain commutations naturally arise between terms that comprise $H_E(t)$ and $U_C(T,t)$. As a result, the first cumulant reduces to integrals over constants and thus,
\begin{eqnarray}
    \mathcal{C}^{(1)}_{O}(T) = \sum_{\mu=M,C}\eta_\mu &&\sum^{p}_{j=1}  g^\mu_j \left( Q_{p:k(j,\mu)}H_\mu Q^\dagger_{p:k(j,\mu)} \right. \nonumber\\
    && -\left. O^{-1} Q_{p:k(j,\mu)}H_\mu Q^\dagger_{p:k(j,\mu)}O\right),\quad
\end{eqnarray}
where $g^M_j = \gamma_j$ and $g^C_j = \beta_j$. The index $k(j,\mu)$ captures the distinctions in the control evolution of Eq.~(\ref{eq:uc-explicit}), where $k(j,M)=j+1$ and $k(j,C)=j$.

\subsection{Stochastic Errors}
We now consider the case of stochastic errors in the variational parameters. Incorporating the zero-mean assumption of the noise model results in $\mathcal{C}^{(1)}_O(T)\equiv 0$; hence, the dynamics are governed by $\mathcal{C}^{(2)}_O(T)$. As will be shown below, the four integrals that comprise the second order term can be exactly calculated for this precision error model.

There are a number of features of our noise model that allow the second cumulant expressions to be conveniently simplified. (1)
The statistical properties of the noise, namely, the lack of cross-correlations, allows for the second cumulant integrals to be partitioned into terms solely proportional to $\tilde{H}_M(t)$ or $\tilde{H}_C(t)$, i.e., $I_j(T)=I_{j,M}(T) + I_{j,C}(T)$. (2) When combined with the piecewise-constant nature of the error Hamiltonian, the integral expressions reduce to a sum of nested integrals only over the domain in which $t_{j-1}\leq s_2\leq s_2\leq t_j$. (3) Leveraging the definition of the ideal QAOA evolution given in Eq.~(\ref{eq:uc-explicit}) and again exploiting commutations between the error Hamiltonian and the ideal propagator, the above expressions once more become integrals over constants. Together, these features result in the following expressions for $I_1(T)$:
\begin{eqnarray}
   I_{1,M}(T)&=&\frac{\Gamma_M}{2}\sum^{p}_{j=1}\beta_j Q_{p:j+1}H^2_M Q^\dagger_{p:j+1}\\
   I_{1,C}(T)&=&\frac{\Gamma_C}{2}\sum^{p}_{j=1}\gamma_j Q_{p:j}H^2_C Q^\dagger_{p:j}.
\end{eqnarray}
Following similar procedures, it can be shown that $I_2(T)=O^{-1}I_1(T)O$, while $I_3(T)=I_{3,M}(T)+I_{3,C}(T)$ with
\begin{widetext}
\begin{eqnarray}
    I_{3,M}(T) &=& \frac{\Gamma_M}{2} \sum^{p}_{j=1}  \beta_j O^{-1} \left(Q_{p\,:j+1} H_{M} Q^\dagger_{p\,:j+1}\right) O \left(Q_{p\,:j+1} H_{M} Q^\dagger_{p\,:j+1}\right),\\
    I_{3,C}(T) &=& \frac{\Gamma_C}{2} \sum^{p}_{j=1} \gamma_j O^{-1} \left(Q_{p\,:j} H_{C} Q^\dagger_{p\,:j}\right) O\left(Q_{p\,:j} H_{C} Q^\dagger_{p\,:j}\right).
\end{eqnarray}
\end{widetext}
It is straightforward to show the remaining term satisfies $I_4(T)=I_3(T)$ as a result of the noise correlation function.

\section{Bounds}
\label{sec:bounds}
Perturbative expansions, like Dyson and Magnus, have proven to be useful tools for developing bounds on open quantum system dynamics~\cite{lidar2008:bounds} . Such bounds have been utilized to investigate and evaluate control schemes designed to mitigate unwanted environmental interactions~\cite{khodjasteh2008:bounds, uhrig2010:bounds,ng2011:dd}. Naturally, the cumulant expansion affords a similar capability that we will exploit here to bound various metrics relevant to QAOA.

\subsection{Error in Expectation Values}
\label{subsec:err-ev}
The expectation value of an observable is a key metric of QAOA that is utilized for both parameter training and calculating performance guarantees. Using the cumulant expansion, we bound the expectation value of an observable $O$ in the presence of a faulty QAOA evolution as follows: first, we find that
\begin{eqnarray}
    |\overline{\braket{O(T)}}| &=& |\text{Tr}[\Lambda(T) \rho_{0}(T) O]|\nonumber\\ 
    &\leq& \|\Lambda(T)\|_\infty\,\, \|O \rho_{0}(T)\|_1.
    \label{eq:bound-on-ev}
\end{eqnarray}
This follows directly from $|\Tr{(AB)}|\leq\|A\|_\infty \|B^\dagger\|_1$, where $\|B\|_1=|\Tr{(B)}|=\sum_i s_i(B)$ is the trace norm of $B$, or equivalently, the sum of singular values $s_i(B)$ \cite{lidar2008:bounds}. Using sub-multiplicativity and the triangle inequality, it is straightforward to show that the operator norm of the error operator can be bounded in terms of the cumulants as
\begin{equation}
   \|\Lambda(T)\|_\infty = \|e^{\mathcal{C}_O(T)}\|_\infty \leq e^{\|\mathcal{C}_O(T)\|_\infty};
   \label{eq:L-bound}
\end{equation}
see Appendix~\ref{app:derivations} for further details. As a result, we obtain the upper bound
\begin{equation}
    |\braket{O(T)}| \leq e^{\|\mathcal{C}_O(T)\|_\infty} \|O \rho_{0}(T)\|_1
\end{equation}
on any observable subject to a faulty QAOA. Note that this bound is independent of the noise model and therefore, holds for both precision errors and more generic spatio-temporally correlated noise models.

Employing a similar analysis, one can derive a bound on the absolute error between the ideal and faulty expectation values. Consider the quantity
\begin{equation}
    |\overline{\Delta\braket{O(T)}}| = |\overline{\braket{O(T)}}-\braket{O(T)}_0|
    \label{eq:abs-err}
\end{equation}
containing the expectation value of $O$ subject to $U(T)$, $\braket{O(T)}$, and the ideal case $\braket{O(T)}_0$ time evolved by $U_0(T)$. In the latter, the error dynamics are equivalent to the identity operator, i.e., $\Lambda(T)\equiv I$. Using this fact, along with properties of the trace, it can be shown that
\begin{eqnarray}
    |\overline{\Delta\braket{O(T)}}| &=& |\Tr{\left[\left(\Lambda(T)-I\right)\rho_{0}(T)O\right]}|\nonumber\\
    &\leq& \left(e^{\|\mathcal{C}_O(T)\|_\infty}-1\right) \|O \rho_{0}(T)\|_1.
    \label{eq:abs-error-ev}
\end{eqnarray}
Additionally, if $\|\mathcal{C}_O(T)\|_\infty\leq 1$, then the inequality $e^x-1\leq (e-1) x$ leads to $|\overline{\Delta\braket{O(T)}}|\leq \|\mathcal{C}_O(T)\|_\infty \|O \rho_{0}(T)\|_1$. Note that in both cases, the error between the ideal and faulty expectation values is dependent upon the distance between $\|\mathcal{C}_O(T)\|_\infty$ and zero. 

In addition to the absolute error, we provide bounds on the mean squared error (MSE). Representing the second moment in the error, the MSE measures the quality of an estimator in terms of both the variance and bias. We define the estimator as $\braket{O(T)}$ and the MSE as
\begin{eqnarray}
    \overline{{\rm MSE}(O(T))} &=& \overline{\braket{\left(O(T) - \braket{O(T)}_0\right)^2}}\nonumber \\
    &=&\overline{{\rm Var}(O(T))} + \left(\overline{\Delta\braket{O(T)}}\right)^2
    \label{eq:mse}
\end{eqnarray}
where ${\rm Var}(O(T))=\braket{O^2(T)} - \braket{O(T)}^2$ is the variance in the faulty observable expectation value. Note that defining the estimator in this particulary way will result in a bound on the MSE in the asymptotic regime when the number of samples $N_S\rightarrow \infty$. Finite sampling effects are further discussed in the supplement.

We derive two bounds on the MSE: one specific to constant coherent errors and another, more general bound characterized by the operator norm of $\Lambda(T)$. The former utilizes formal integration of $\tilde{U}_E(t)$ -- similar to typical Trotter error bounding -- to upper bound the MSE in terms of $\|H_E(t)\|_\infty$. The latter employs the techniques used in Eq.~(\ref{eq:bound-on-ev}) and applies to any spatio-temporally correlated, weak-noise model. The second approach inherently provides a bound for constant coherent errors as well, however, we find that the first approach yields a tighter bound than its more general counterpart for this particular error model.

In the case of coherent errors, the MSE is bounded as
\begin{equation}
    {\rm MSE}(O(T))\leq 2\|O\|^2_\infty \left(1 + 2 h^2_E\right),
    \label{eq:mse-bound-const}
\end{equation}
where
\begin{equation}
    h_E = \theta_M(T) \|H_M\|_\infty + \theta_C(T) \|H_C\|_\infty
    \label{eq:he}
\end{equation}
with $\theta_{M}(T)=\int^{T}_0 dt |\eta_M(t)| [1-s(t)]$ and $\theta_{C}(T)=\int^{T}_0 dt |\eta_C(t)| s(t)$. This result follows from a bound on the variance ${\rm Var(O(T))} \leq 2\|O\|^2_\infty$ and on the bias  $\Delta \braket{O(T)})\leq\int^{T}_0 \|H_E(t)\|_\infty dt\leq h_E$.

More generally, the variance can be bounded in terms of the error operator according to
\begin{eqnarray}
    \overline{{\rm Var(O(T))}} \leq \|\Lambda(T)\|_\infty\, &&\left(\|\Lambda(T)\|_\infty\, \|O\rho_0(T)\|^2_1 \right. \nonumber\\
    &&\left.+ \|O^2 \rho_0(T)\|_1\right).
    \label{eq:var-bound}
\end{eqnarray}
This follows directly from an application of the triangle inequality and the variant of H\"{o}lder's inequality used in Eq.~(\ref{eq:bound-on-ev}). A bound on the bias [the second term in Eq.~(\ref{eq:mse})] is obtained in a similar manner to that of Eq.~(\ref{eq:abs-error-ev}), namely,
\begin{equation}
    \left(\overline{\Delta\braket{O(T)}}\right)^2 \leq \|\Lambda(T)-1\|^2_\infty \, \|O\rho_0(T)\|^2_1.
    \label{eq:bias-bound}
\end{equation}
The MSE achieves a quadratic scaling in the norm of the error operator and therefore, the cumulant sum. This can be observed by incorporating Eq.~(\ref{eq:L-bound}) to explicitly express the MSE in terms of the norm of the cumulant sum. If $\|\mathcal{C}_O(T)\|_\infty \leq 1$ is assumed, the MSE scales as $O(\|\mathcal{C}_O(T)\|^2_\infty)$, explicitly conveying a quadratic scaling.

\subsection{Approximation Ratio}
The bound derived in Eq.~(\ref{eq:abs-error-ev}) naturally leads to a bound on the absolute error in the approximation ratio. Specifically, we aim to bound the difference between the faulty and ideal approximation ratios for evolution dictated by the optimal \emph{ideal} variational parameters. We show that the absolute error can be expressed in terms of the operator norm of the cumulant series $\mathcal{C}_O(T)$. As a result, one may utilize this bound to quantify the effect of various types of precision errors on the approximation ratio. While we focus on a general bound here, subsequent sections consider specific canonical QAOA problems.

In the ideal scenario, an $\epsilon^*$-approximate QAOA achieves a performance guarantee with the variational parameters $(\bm{\gamma}^*, \bm{\beta}^*)$, or equivalently a total algorithmic runtime $T^*$. The optimal approximation ratio $\epsilon^*=F(\bm{\gamma}^*, \bm{\beta}^*)/C_{\rm max}$ is expressed in terms of the average energy with respect to the problem Hamiltonian, $F(\bm{\gamma}^*,\bm{\beta}^*)$, i.e., the typical QAOA objective function. This quantity is synonymous with Eq.~(\ref{eq:avg-o}) rescaled by $C_\mathrm{max}$, where $O=H_C$ and the evolution has been evaluated at the optimal variational parameters.

In order to evaluate the error induced by $H_E(t)$, and more specifically precision errors, consider the average energy resulting from the faulty dynamics $\widetilde{F}(\bm{\gamma}^*, \bm{\beta}^*)$. As can be anticipated, the average energy can be associated with a faulty approximation ratio $\widetilde{\epsilon^*}=\widetilde{F}(\bm{\gamma}^*, \bm{\beta}^*)/C_{\rm max}$. The upper bound on the absolute error $|\Delta \epsilon^*| = |\widetilde{\epsilon^*} - \epsilon^*|$ is
\begin{eqnarray}
    |\overline{\Delta \epsilon^*}| &=& \frac{1}{C_{\rm max}}|\widetilde{F}(\bm{\gamma}^*, \bm{\beta}^*) - F(\bm{\gamma}^*, \bm{\beta}^*)| \nonumber\\
    &\leq& \frac{1}{C_{\rm max}} \|\Lambda(T^*)-1\|_\infty\,\, \|H_C\, \rho_{0}(T^*)\|_1\nonumber\\
    &\leq& \frac{1}{C_{\rm max}} \left(e^{\|\mathcal{C}_O(T^*)\|_\infty}-1\right) \|H_C\, \rho_{0}(T^*)\|_1,\quad
    \label{eq:approx-bound}
\end{eqnarray}
which follows directly from Eq.~(\ref{eq:abs-error-ev}). In Sec.~\ref{subsec:cum-bounds}, we further bound $\|\mathcal{C}_O(T)\|_\infty$ for the error models considered in Sec.~\ref{sec:err-an}, i.e., constant coherent errors and stochastic errors. These bounds are incorporated into Eq.~(\ref{eq:approx-bound}) in Sec.~\ref{sec:examples}, where specific QAOA problems are investigated.


\subsection{Error in Training}
\label{subsec:err-training}
QAOA variational parameter training routinely involves a classical gradient descent algorithm. Gradients are either analytically determined or approximated via a finite difference approximation. In either case, parameters are updated based on estimates of the average energy using states time evolved by the QAOA with variational parameters acquired from previous iterations. Thus, the expectation value plays an important role in the effective training of the QAOA. When presented with a faulty QAOA, subject to systematic or environment noise sources, errors induced in the expectation value propagate to the gradient, and ultimately, impact parameter training. 

Here, we aim to evaluate the impact of errors on the gradient by bounding the absolute error in the gradient of the expectation value. First, we focus on open-loop (offline) optimization protocols that utilize exact gradient expressions for parameter training. We define the error between the faulty and ideal gradients for a single variational parameter
\begin{equation}
    |\partial_{\gamma_j} \overline{\Delta \braket{O(T)}}| = |\partial_{\gamma_j}\overline{\braket{O(T)}} - \partial_{\gamma_j}\braket{O(T)}_0|.
\end{equation}
Applying the triangle inequality, the error can be separated into two terms as follows:
\begin{eqnarray}
    |\partial_{\gamma_j}\overline{\Delta \braket{O(T)}}| 
    &\leq&|\Tr{\left[\left(\partial_{\gamma_j}\Lambda(T)\right)\rho_{0}(T) O\right]}|\nonumber\\
    & & + |\Tr{\left[ \left(\Lambda(T)-1\right) \partial_{\gamma_j}\rho_{0}(T) O\right]}|.
    \label{eq:grad-partial}
\end{eqnarray}
The first term is characterized by the differentiation of the error operator, while the latter includes a differentiation of the ideal time-evolved density operator. We bound each term individually, starting with
\begin{eqnarray}
    |\Tr{\left[\left(\partial_{\gamma_j}\Lambda(T)\right)\rho_{0}(T) O\right]}| &\leq& \|\partial_{\gamma_j}\Lambda(T)\|_\infty\,\, \|O \rho_{0}(T)\|_1 \nonumber \\
&\leq& \lambda(T) \,\,\|O \rho_{0}(T)\|_1.
\label{eq:grad-1}
\end{eqnarray}
The first inequality follows from an application of H\"older's inequality, while the second incorporates
\begin{eqnarray}
    \|\partial_{\gamma_j}\Lambda(T)\|_\infty &\leq&  e^{\|\mathcal{C}(T)\|_\infty} \frac{e^{2\|\mathcal{C}(T)\|_\infty}-1}{2\|\mathcal{C}(T)\|_\infty} \|\partial_{\gamma_j}\mathcal{C}(T)\|_\infty\nonumber\\
    &=& \lambda(T)
    \label{eq:lambda}
\end{eqnarray}
Briefly, this bound is obtained by expressing $\Lambda(T)$ as an exponential of the cumulant series and using the general definition of a derivative of an exponentiated operator. 

The second term in Eq.~(\ref{eq:grad-partial}) can be bounded via H\"older's inequality and further simplified using the Liouville-Von Neumann equation. Namely, we find
\begin{eqnarray}
    |\Tr{\left[ \left(\Lambda(T)-1\right) \partial_{\gamma_j}\rho_{0}(T) O\right]}| &\leq& \|\Lambda(T)-1\|_\infty \nonumber\\
    & & \times \|\tilde{O}(T_j, T) \left[H_C,\rho_0(T_j)\right]\|_1, \nonumber \\
    \label{eq:grad-2}
\end{eqnarray}
where $\tilde{O}(T_j, T)=Q_{p:j} O Q^\dagger_{p:j}$. Note that the second term in Eq.~(\ref{eq:grad-2}) follows from observing the connection between $\|O^\dagger \partial_{\gamma_j}\rho_0(T)\|_1$ and the dynamical equation for the density matrix up to time $T_j$, i.e., $\partial_{\gamma_j}\rho_{0}(T) = -i Q_{p:j} \left[H_C,\rho_0(T_j)\right] Q^\dagger_{p:j}$. The supplement contains further elaboration on this bound and the others shown above.

Commonly, QAOA training leverages approximate gradient expressions in place of exact gradients. This most notably occurs when closed loop optimization protocols employ some degree of hardware-in-the-loop querying to train variational parameters. First order gradients are calculated via finite difference approximations and subsequently incorporated into stochastic gradient descent protocols, such as simultaneous perturbation stochastic approximation (SPSA)~\cite{spall1992:spsa}.

The training error in stochastic optimization protocols can be associated with the quality of the estimator, and thus, the MSE. In the asymptotic limit, this training error can be bounded by the expressions given in Sec.~\ref{subsec:err-ev}. Coherent errors yield an upper bound of
\begin{equation}
    {\rm MSE}(O(T))\leq 2\|H_C\|^2_\infty \left(1 + 2 h^2_E\right),
    \label{eq:mse-coherent}
\end{equation}
on the training error, while the sum of Eqs.~(\ref{eq:var-bound}) and (\ref{eq:bias-bound}) with $O=H_C$ specifies a general bound in terms of the error operator. Finite sampling effects are pertinent to approximate gradient-based algorithms. Contributing additional corrections to the above bounds, such effects are elaborated upon in the supplement.


\subsection{Distance Between Unitaries}
Error measures expressed in terms of expectation values provide insight into the deleterious effects of noisy QAOA evolution with respect to the measurement of an observable. Although such measures are of practical importance for QAOA, from a theoretical perspective, it can be constructive to eliminate state- and observable-dependence and focus solely on the unitary propagator. Such distance measures are commonly used in optimized control~\cite{wilhelm2020:qc} and Trotter error analysis~\cite{childs2021:te} to achieve a holistic understanding of error propagation. 

In an effort to obtain such an understanding for the precision error problem in QAOA, we examine the distance between the faulty QAOA and ideal QAOA evolution
\begin{eqnarray}
   \Delta U = \|U(T) - U_{0}(T)\|_2.
   \label{eq:delta-u}
\end{eqnarray}
Note that $\|A\|_2=\sqrt{\Tr{A^\dagger A}}$ defines the Frobenius norm. We bound this quantity, focusing particularly on the ensemble average-squared distance, i.e.,
\begin{eqnarray}
    \overline{(\Delta U)^2} = \Tr{\left[2I-\overline{\tilde{U}_E(T)} - \overline{\tilde{U}^\dagger_E(T)}\right]}.
\end{eqnarray}
The above expression includes a noise averaged unitary that differs from that introduced in Eq.~(\ref{eq:lambda-cum}). Namely, the unitary is not generated by the effective Hamiltonian $\tilde{H}_{E,O}(t)$, but rather $\tilde{H}_E(t)$. This does not create any issues, however, as a cumulant expansion 
\begin{eqnarray}
    \overline{\tilde{U}_E(T)} = \overline{\mathcal{T}_+ e^{-i\int^{T}_0 \tilde{H}_E(t) dt}} 
    = e^{\mathcal{C}(T)}
\end{eqnarray}
can still be obtained, with $\mathcal{C}(T)=\sum^\infty_{k=1}(-i)^k\mathcal{C}^{(k)}(T)/k!$ denoting the cumulant terms rewritten in terms of $\tilde{H}_E(t)$. Incorporating the expansion into Eq.~(\ref{eq:delta-u}) leads to
\begin{eqnarray}
   \overline{(\Delta U)^2} &=& \Tr{\left[2I-e^{\mathcal{C}(T)} - e^{\mathcal{C}^\dagger(T)}\right]}\nonumber\\
   &\leq& 2 \left|\Tr{\left(I-e^{\mathcal{C}(T)}\right)}\right| \nonumber\\
   &\leq& 2 \left\|I-e^{\mathcal{C}(T)}\right\|_\infty \nonumber\\
   &\leq& 2\left(e^{\|\mathcal{C}(T)\|_\infty}-1\right).
\end{eqnarray}
This bound possesses similar qualities to the bound obtained for the absolute error in expectation values, with the primary distinction being the presence of an observable-\emph{independent} cumulant expansion.

\section{Numerical Simulations and Empirical Bounds}
\label{sec:examples}
In this section, we discuss the effect of precision errors on two specific examples of QAOA circuit implementations. In the first example, we examine the Grover's Search Algorithm using QAOA~\cite{Jiang2017}  and second we  consider obtaining the ground state of the nearest-neighbor Ising model on a ring~\cite{Ho_2019}, which also can be mapped to the  2-SAT problem on a ring~\cite{Farhi2000}. While the Grover algorithm provides an example of a QAOA algorithm with an analytical solution for the optimal circuit depth and angles, the circuit depth increases exponentially with the number of qubits. In the second example, we obtain the ground state of the Ising problem using a number of QAOA layers that scales linearly in the number of qubits.

We numerically simulate precision errors and investigate its effects on the accuracy of the implementation of the algorithm. The errors $\eta_{C,M}$ modify the QAOA phases, $\gamma\rightarrow \gamma(1+\eta_C)$ and $\beta\rightarrow \beta(1+\eta_M)$; see Eq.~(\ref{eq:errors}) for definitions of $\eta_\nu$. These errors are drawn from a normal distribution: $\eta_{\nu} \in \mathcal{N}(\eta,\sqrt{\Gamma})$ with $\eta$ being the mean, $\Gamma$ the variance of the normal distribution and $\nu \equiv M,C$. The stochastic error case reduces to $\eta_{\nu} \in \mathcal{N}(0,\sqrt{\Gamma})$ [$\Gamma_{\nu}(t-t^\prime)=\Gamma \delta(t-t^\prime)$, i.e., no correlation between subsequent errors in $\gamma_k$ and $\beta_k$] and the coherent error cases reduces to $\eta_{\nu} \in \mathcal{N}(\eta,0)$ [$\Gamma_{\nu}(t-t^\prime)=\eta^2$, i.e., constant errors in all $\gamma_k$ and $\beta_k$]. We average our results over a 1000 noise realizations for the case of the stochastic error, and 1 realization for the coherent case.

 In order to numerically extract the loss in accuracy of the computation with 
 increasing precision error, we model the noise-averaged expectation value using a decay function dependent on the number of layers and noise parameters,
\begin{align}
    \overline{\braket{H_C}}\approx \braket{H_C}_0 \mathcal{S}_{H}(p,\mu,\sigma)
\end{align}
where $\braket{H_C}_0$ corresponds to the noiseless expectation value . The analytical form of the decay function, $\mathcal{S}_{H}(p)$ depends on the nature of the precision errors (stochastic vs coherent) and the number of layers. In the following, we show that the expectation value decays exponentially away from the noiseless value, and this behavior is generic.

\subsection{Grover's Search Algorithm}\label{subsec:Grover}
The Grover search problem involves finding a marked element $\ket{m}$ in a unsorted database consisting of $N=2^n$ elements, where $n$ is the number of qubits required to represent the database. For a marked state of $\ket{m}=\ket{0\cdots 0}$, we have the following phase and mixer Hamiltonians, 
\begin{align}
H_{\rm C} &= |0\cdots0\rangle\langle 0\cdots0|=\bigotimes_{i=1}^n\frac{I+Z_i}{2}\\
H_M &=\sum_{i=1}^n X_i
\label{eq:x-mixer}
\end{align}
Clearly, the phase Hamiltonian corresponds to a fully connected Ising Hamiltonian. Unlike the cost function for typical optimization problems, the cost function has all possible tensor product of $Z_i$ operators. However, note that $||H_{\rm C}||=1$. Also, note that the Hamiltonian encoding the cost function, $H_{\rm C}$, while clearly hermitian, is non-invertible.

\begin{figure}
\includegraphics[width=\linewidth]{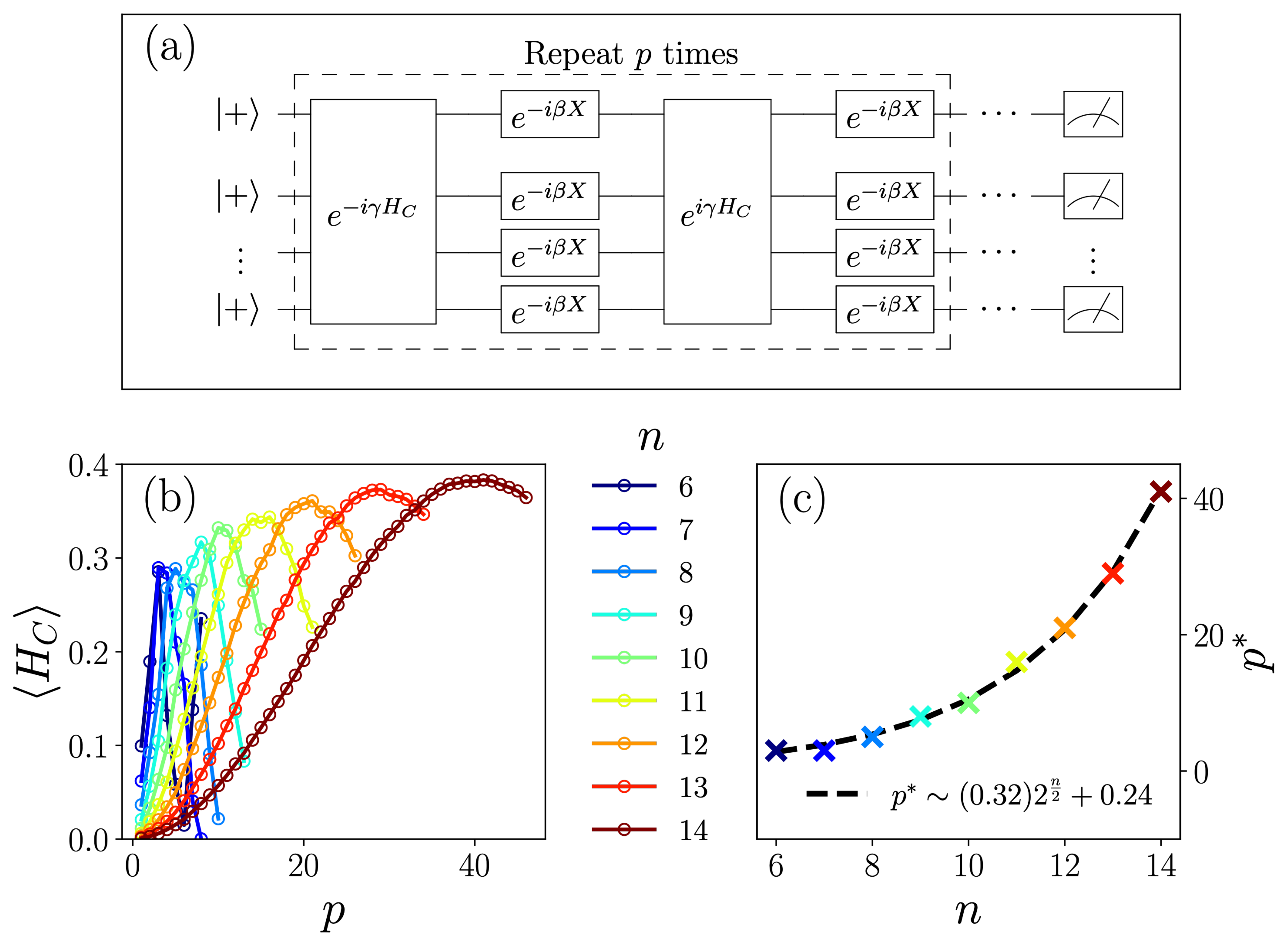}
\caption{Noisless implementation of the Grover search algorithm. (a) The circuit for implementing the Grover Search algorithm using QAOA~\cite{Jiang2017}. (b) The expectation value  of the cost function as a function of the number of layers $p$. (c) The optimal number of layers, $p^*\sim 2^{n/2}$ that  maximizes the cost function increases proportionally with the problem size. We extract the functional form of the optimal number layers, $p^*\approx (0.32)2^{n/2}+0.24$. }
\label{fig:Grover-QAOAcircuit}
\end{figure}

Typically, for the Grover search problem, each layer of the QAOA ansatz has the following structure~\cite{Jiang2017},
\begin{align}
|\psi_{\rm out}\rangle &=\left[W(\gamma,\beta)\right]^p\ket{+}^{\otimes n}\\
&=\prod_{k=1}^{2p}\left[ e^{-i\beta \sum_j X_j} e^{(-1)^{k+1}i\gamma H_C}\right] \ket{+}^{\otimes n}
\label{eq:Grover-circuit}
\end{align}
with $\gamma=\pi$ and $\beta=\frac{\pi}{n}$.  The QAOA algorithm applies the unitary $W(\gamma,\beta)$  $p$ times. The quantum circuit for a single layer, is shown in Fig.~\ref{fig:Grover-QAOAcircuit}(a). We simulate the Grover QAOA circuits using the built-in simulator in Cirq~\cite{cirq_2020}. The expectation value of the cost function, which corresponds to the probability of measuring the marked state, oscillates as a function of the number of layers $p$. In Fig.~\ref{fig:Grover-QAOAcircuit}(b) we show this probability as a function of $p$. We obtain an expression for the optimal number of layers for Grover search as a function of the $n$ by numerically fitting the first maximum in $\langle H_C\rangle$,
\begin{align}
    p^*(n) \approx a_1 2^{n/2}+a_2\label{eq:popt-grover}
\end{align}
$a_1=0.32$, $a_2=0.24$; see Fig.~\ref{fig:Grover-QAOAcircuit} (b). This scaling of the optimal  number of layers for Grover is in line with the theoretical analysis in Ref.~\cite{Jiang2017}, namely $p^*\sim O(\sqrt{N})$.
\begin{figure}
    \centering
    \includegraphics[width=\linewidth]{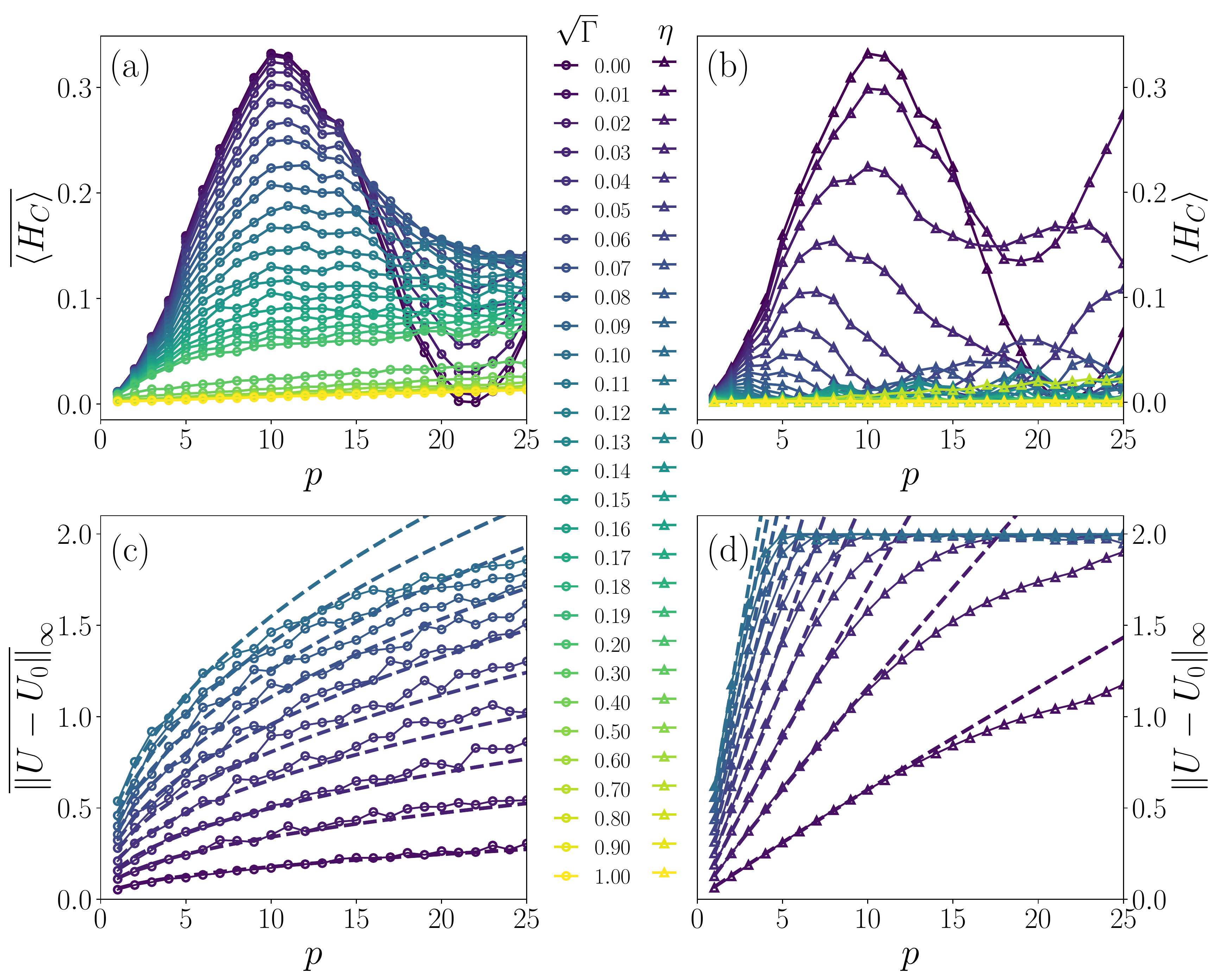}
    \caption{Performance of Grover QAOA circuits as a function of number of layers $p$ in the presence of precision errors or multiplicative control noise. We consider the effect of two types of precision errors, (i) [Left Column] Stochastic errors, drawn from a normal distribution $\mathcal{N}(0,\sigma)$; and (ii) [Right Column] Coherent errors, drawn from a normal distribution $\mathcal{N}(\mu ,0)$. The top row shows the effect of the precision errors on the calculated expectation value, $\overline{\langle H_C\rangle}$ where $\overline{\cdots}$ indicates noise averaging. The bottom row shows the average distance between the the faulty and the perfect QAOA  unitary as characterized by the average difference in the $\infty$-norm, $\overline{\|U-U_0\|}_\infty$. The dashed line shows the behavior of a scaling function, $\mathcal{S}_{U}(\sigma,\mu,p)$ that describes well the behavior of the increasing separation for stochastic and coherent errors, the numerical parameters are obtained in  the appendix. The simulations are done for $n=10$ qubits, with  $1000$ noise-realizations for the stochastic case and $1$ realization for the coherent error case. }
    \label{fig:grover1}
\end{figure}

The Grover cost function, $H_C$ is not invertible, which poses an issue for the cumulant expansion and the definition of the error operator $\Lambda(T)$ [Eq.~(\ref{eq:avg_O_lamb})].  However, we can be rewrite $H_C$ as,
\begin{align}
    H_C&=\frac{1}{N}I + O,\label{eq:grover-hc-sum}\\
    \textrm{where, } O&=\frac{1}{N}\sum_{\{A_i\}} A_{1} \otimes \cdots A_{n}
\end{align}
where, $A_i\in \{I,Z_i\}$ and $O$ is invertible and traceless. Using this representation of $H_C$, we can  employ the bounds derived in the previous sections to bound the expectation value of $H_C$. We consider the effects of stochastic and coherent noise on the measured expectation value of the cost Hamiltonian. We expect a noise-averaged expectation value of,
\begin{align}
    \overline{\braket{H_C}}&=\frac{1}{N}I + \overline{\braket{O}}=\frac{1}{N}I + \braket{O}_0\mathcal{S}_O(p^*)\\
    &=\frac{1}{N}\left[1-\mathcal{S}_O(p^*)\right]+\braket{H_C}_0 \mathcal{S}_O(p^*),
\end{align}
where $\mathcal{S}_O(p^*)$ is the decay function specific to the operator $O$. 

\begin{figure}
    \centering
    \includegraphics[width=\linewidth]{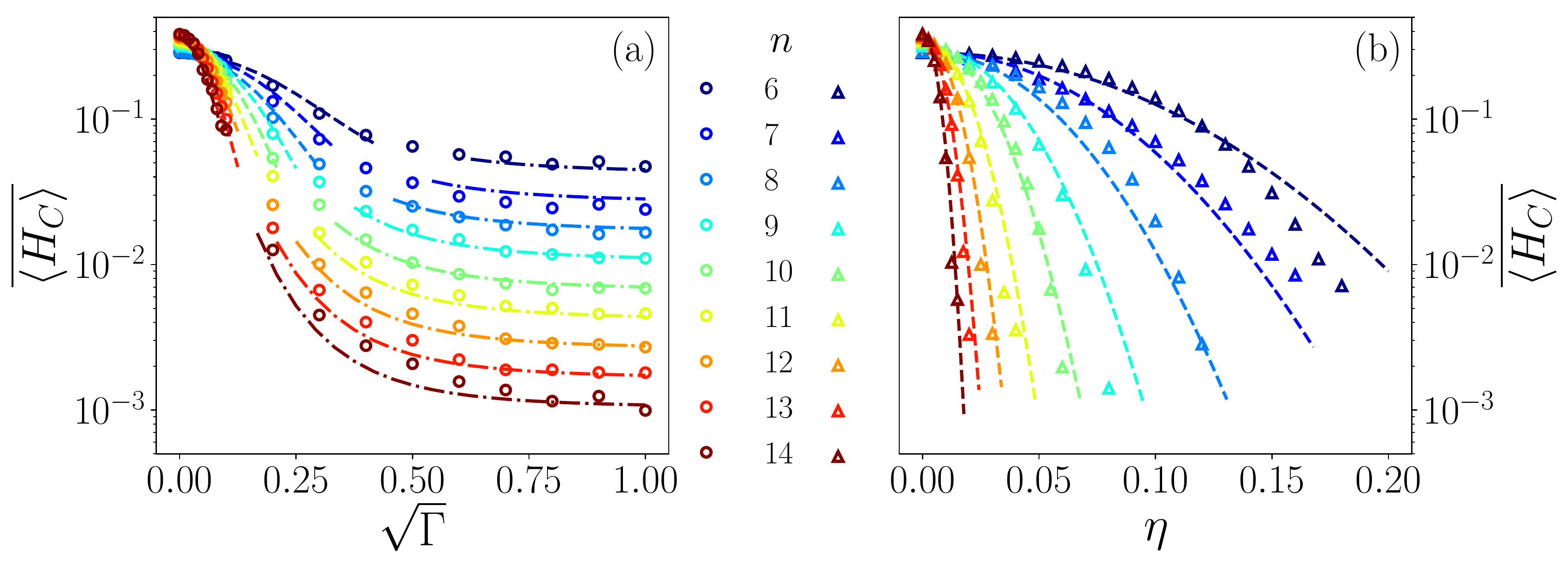}
    \caption{Performance of Grover QAOA algorithm for the optimal number of layers ($p^*$) as a function of noise-strength [(a) $\sigma$ or (b) $\mu$] and system size (color). (a) Noise-averaged expectation value, $\langle H_C\rangle$ as a function of the strength of the stochastic error (open circles). The lines (dashed and dashed-dot) indicate the behavior from the scaling function. (b) Expectation value as a function of strength of coherent error (open triangles) and the corresponding scaling function (dashed line).}
    \label{fig:grover-size-dependence}
\end{figure}

In the limit of weak noise, we fit the measured  expectation value to the following phenomenological decay function,
\begin{align}
    \textrm{Weak noise limit:} \overline{\braket{H_C}}&\approx \braket{H_C}_0 \mathcal{S}_{O}(p^*)
\end{align}
In Figs.~\ref{fig:grover1} and~\ref{fig:grover-size-dependence}, we numerically study the effects of precision error. Fig.~\ref{fig:grover1} (a) and (b), we study the role of stochastic~($\eta=0$, $\Gamma\neq 0$) and coherent errors ~($\eta\neq0$, $\Gamma= 0$) as a function of number of layers $p$ with the weak noise limit corresponding to $\sigma \ll 1$ and $\mu\ll 1$ respectively. We numerically obtain a fit for the decay function in Fig.~\ref{fig:grover-size-dependence} (a) and (b) in the weak noise limit (shown as dashed lines) for the optimal Grover QAOA algorithm, 
\begin{align}
    \mathcal{S}_{O}(p^*)&\approx \braket{H_C}_0\exp[{-\chi_O(p^*)}]\\
    \textrm{with, } \chi_O(p^*)&=
    \begin{cases} 
    -\Gamma p^*/\Gamma_s, & \textrm{[stochastic error]} \\ 
    -\eta^2 {p^*}^2/\eta_c^2, & \textrm{[coherent error]} \end{cases}
\end{align}
where $\sqrt{\Gamma_s}\approx 0.47$ and $\eta_c\approx 3.46$. Clearly, the calculated expectation value decays exponentially with precision error scale. Note that the implementation of the optimal Grover algorithm requires $p_{\rm opt}$ that grows with the problem size. Conversely, our results imply that for an accurate implementation of the optimal Grover QAOA circuit, the precision errors need to be  bounded from above, $\Gamma\ll\Gamma_s/p^*$ and $\eta \ll \eta_c/p^*$.


\begin{figure}
    \centering
    \includegraphics[width=\linewidth]{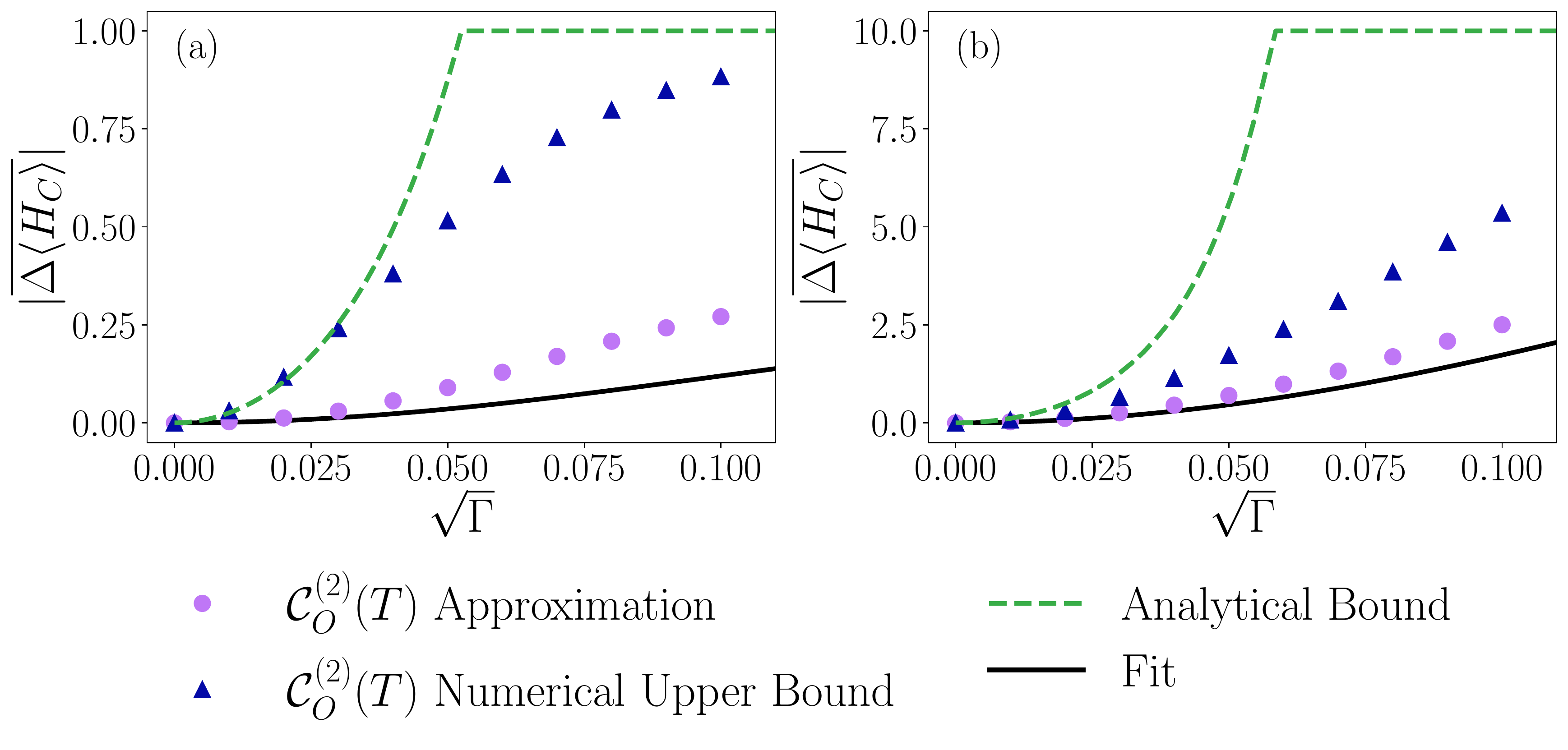}
    \caption{Comparison of the numerically obtained scaling functions, with the cumulant expression and bounds for stochastic precision errors. Second cumulant approximation (purple circles) shows good agreement with scaling functions (black lines) in the weak noise regime. Numerically calculated error bounds using the second cumulant approximation (blue triangles) are shown in addition to analytically calculated cumulant bound (green dashed) of Eq.~(\ref{eq:abs-error-ev}). Plots summarize results for $n=10$ qubits for (a) Grover's Search and (b) GHZ state preparation.}
    \label{fig:grover-bounds}
\end{figure}

 We compare the numerically obtained $\mathcal{S}_{O}(p^*)$ to the decay functions derived via the cumulant expansion and its associated bounds. Specifically, we focus on the absolute error in the expectation value, and compare the numerically obtained absolute error to three different cumulant calculations for the case of stochastic precision errors. First, we study the absolute error using the second cumulant approximation to the error operator. We compare this approximation to both numerically and analytically obtained bounds to assess their efficacy in the Grover's search setting. A summary of the analysis is shown in Fig.~\ref{fig:grover-bounds}.
 
The absolute error is estimated by approximating the expectation value dynamics using the second-cumulant-truncated error operator. The cost Hamiltonian is expressed as a sum of terms [see Eq.~(\ref{eq:grover-hc-sum})], which naturally leads to an expectation value of the form
 \begin{equation}
     \overline{\braket{H_C}}=\sum_{\mathcal{O}\in\{I/N,O\}}\tr{\Lambda_{\mathcal{O}}(T)\rho_S(T)\mathcal{O}}.
     \label{eq:grover-avgO}
     \end{equation}
 By truncating each error operator as $\Lambda_{\mathcal{O}}(T)\approx e^{-C^{(2)}_{\mathcal{O}}(T)/2}$, we obtain an expression for the approximate dynamics of the expectation value of $H_C$. Subsequently, an estimate of the absolute error is obtained by substituting Eq.~(\ref{eq:grover-avgO}) into Eq.~(\ref{eq:abs-err}). 
 
 The approximate dynamics are compared against bounds on the absolute error derived in Section~\ref{subsec:err-ev}. Specifically, we consider a variant of the upper bound given in Eq.~(\ref{eq:abs-error-ev}) for non-invertible observables:
 \begin{equation}
     |\overline{\Delta\braket{H_C}}| \leq \|\sum_{\mathcal{O}\in\{I/N,O\}}\mathcal{O}\Lambda_{\mathcal{O}}(T)- H_C\|_\infty;
 \end{equation}
 see Appendix~\ref{subsec:noninv-o} for further details. We consider a numerical evaluation of this upper bound by truncating the error operator to second order and calculating the bound numerically. In addition, we estimate the bound analytically via
 \begin{eqnarray}
     |\overline{\Delta\braket{H_C}}| &\leq& \sum_{\mathcal{O}\in\{I/N,O\}} \|\mathcal{O}\|_\infty \left(e^{\|C^{(2)}_{\mathcal{O}}(T)\|_\infty/2}-1\right)\nonumber\\
     &\leq& \min\left[2\left(e^{4\pi p^* \Gamma}-1\right),\|H_C\|_{\infty} \right],
 \end{eqnarray}
 where the upper bound on the second cumulant is
 \begin{equation}
    \|C^{(2)}_{\mathcal{O}}(T)\|_\infty\leq 8\pi p^* \Gamma\sim 8\pi \sqrt{N} \Gamma.
\end{equation}
 The bound on the $\|C^{(2)}_{\mathcal{O}}(T)\|_\infty$ holds for each constituent observable $\mathcal{O}$ defined for Grover's search. Note that the upper bound expression is clipped at a maximum value of $\|H_C\|_{\infty}$ as this represents the physical bound on possible QAOA solutions.


The results of the cumulant comparison are summarized in Fig.~\ref{fig:grover-bounds}(a) for Grover's Search. The second cumulant approximation (purple circles) agrees well the numerically obtained fit in the weak noise regime.  Analytical bounds (green dashed lines) provide a modest representation of precision error-dependence, deviating more substantially as the variance increases. Numerical bounds (blue triangles) and approximate dynamics follow similar trends, indicating that as the variance increases, higher order cumulants likely contribute more significantly to the dynamics.

In addition, we evaluate the effect of precision errors by calculating the average distance between the faulty and ideal unitary.  Our results are summarized in   in Fig.~\ref{fig:grover1} (c) and (d) which show the stochastic and coherent error case, respectively.  In the limit of weak noise, the average distance for 
\begin{align}
    \overline{\|U-U_0\|}_\infty &\propto
    \begin{cases}
      \sqrt{\Gamma p},& \textrm{[stochastic error]} \\ 
     \eta p, &\textrm{[coherent error]} \\ 
    \end{cases}
\end{align}
The average distance between the noisy and noiseless unitary grows with the number of layers of the QAOA circuit.

Finally, let us discuss the limit of large stochastic noise, $\Gamma p^*\gg 1$. Assuming that the output of a noisy QAOA evolution is a completely random state, one would expect that $\overline{\braket{H_C}}\sim 1/N=1/2^{n/2}$. However, in this limit, the noisy-QAOA does not give rise to uniformly random states. Instead, we observe saturation in the expectation value that scales with problem size according to
\begin{align}
    \braket{H}_{\rm sat}\propto 2^{\xi n},
\end{align}
where $\xi\approx 0.67$. The approach to the saturation value is governed by a power-law behavior,
\begin{align}
    \overline{\braket{H_C}}-\braket{H_C}_{\rm sat}\propto \frac{1}{(\Gamma p^*)^\alpha},
\end{align}
where $\alpha\approx 1.61$. The details for the fit to the various parameters, $\sigma_s$, $\mu_c$, $\xi$ and $\alpha$ are provided in  Appendix~\ref{app:numerics}.

\subsection{Ising Instances}
\label{subsec:IsingInstances}



Essentially all combinatorial optimization problems may be cast as two-body Ising Hamiltonians~\cite{Barahona1982,lucas} of the form $H=\sum_{ij} J_{ij} Z_i Z_j + \sum_i h_i Z_i$. 
It is therefore natural to study the effects of finite precision errors on problems of this type. A paradigmatic model of the above form, which is therefore convenient to analyze, is the one-dimensional, nearest-neighbor Ising model. 
The problem Hamiltonian of this model for $n$ qubits, is given by 
\begin{equation}
    H_C = \sum_{i=1}^{n} Z_i Z_{i+1},
    \label{eq:ghz-hc}
\end{equation}
where the summation is taken over neighboring spins on a 1D chain with periodic boundary conditions, i.e., $Z_{n+1}\equiv Z_1$. For instance, this particular cost function can be used to find the solutions for 2-SAT on a ring~\cite{Farhi2000}. 

The cost function Hamiltonian for this Ising model is maximized by a state that is any superposition of all qubits in the $\ket{0}$ or $\ket{1}$ state. In fact, by restricting to a given parity sector satisfying $\prod_{i=1}^{n} X_i=1$, 
this cost function can be used in the QAOA setting as an algorithm for preparing a GHZ state ~\cite{Ho_2019,pagano2020:qaoa},
\begin{align}
    \ket{GHZ}&=\frac{1}{\sqrt{2}}\left(\ket{0}^{\otimes n}+\ket{1}^{\otimes n}\right).
\end{align}
We follow Ref.~\cite{Ho_2019} to use the QAOA ansatz for preparing the GHZ state. Note that we have flipped the sign of the cost Hamiltonian compared with Ref.~\cite{Ho_2019} as we have recast the objective from minimization to maximization. A similar procedure is utilized for the mixer Hamiltonian as well; thus, $H_M$ is defined by Eq.~(\ref{eq:x-mixer}). It was shown that the optimal number of layers to prepare a GHZ  state scales polynomially with the number of qubits. In particular, $p^*=n/2$, with the following structure for the variational circuit ansatz:
\begin{align}
    U_k(\gamma_k,\beta_k)=e^{-i\beta_k \sum_j X_j }e^{-i\gamma_k H_C}, \\
    \ket{GHZ}=\left[\prod_{k=1}^{n/2}U_k(\gamma_k,\beta_k)\right] \ket{+}^{\otimes n}.
\end{align}
The optimal parameters to generate GHZ states for different system sizes are taken from Ref.~\cite{Ho_2019} (Appendix A) to test the susceptibility  of the state  preparation to precision errors.

\begin{figure}
    \centering
    \includegraphics[width=\linewidth]{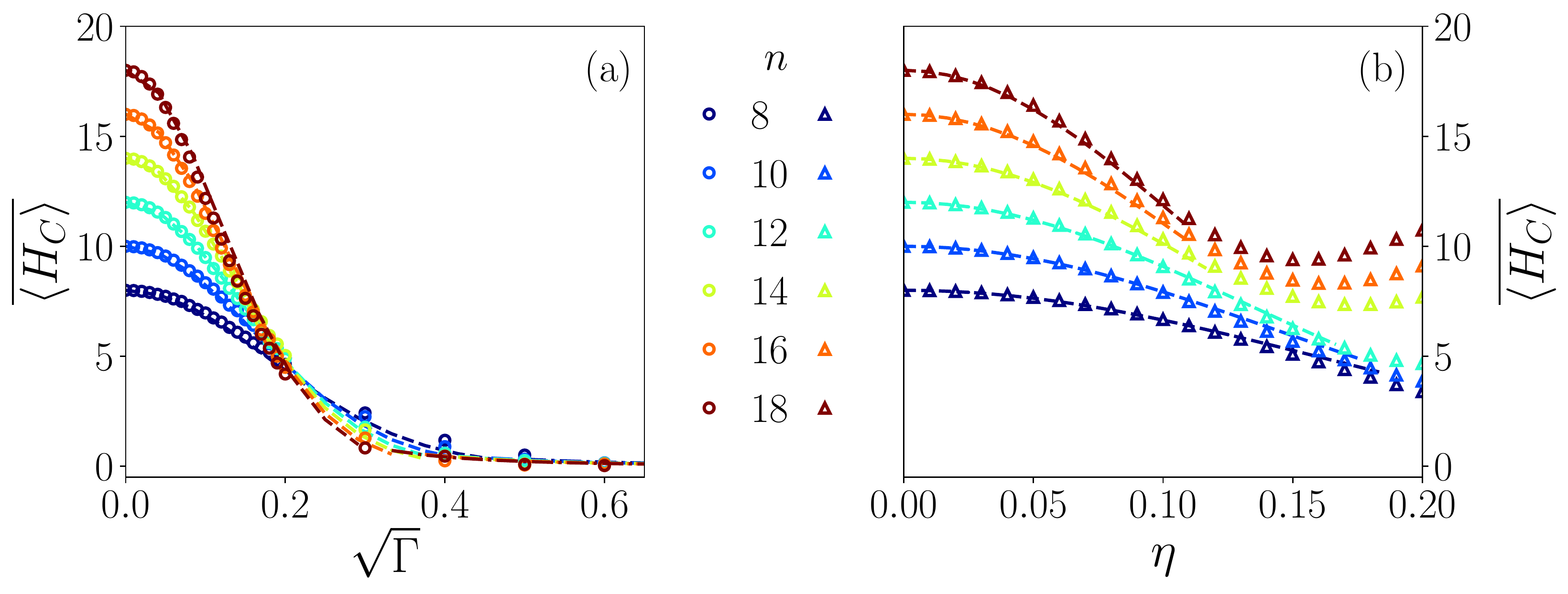}
    \caption{Preparation of the GHZ state using the QAOA algorithm with the optimal number of layers ($p=p^*$) as a function of noise strength[(a) $\sigma$ or (b) $\mu$] corresponding to noise generated from a normal distribution $\mathcal{N}(\mu,\sigma)$. (a) Average expectation value (open circles) of the cost function for the case of stochastic error ($\mu=0$), the lines (dashed,dash-dot) indicate the scaling fit to the data. (b) Expectation value (open triangles) as a function of the strength of coherent error ($\sigma=0$) and the corresponding scaling fit shown as a dashed line.}
    \label{fig:ghz_size_dependence}
\end{figure}

We numerically model the noise-averaged expectation value as a function of the noise strength and determine scaling fits for the GHZ state preparation problem. A summary of the results are shown in Fig.~\ref{fig:ghz_size_dependence} for both stochastic and constant precision errors. Numerical simulations are performed using 1000 realizations for stochastic errors and one realization for the coherent error case, using code adapted from Refs.~\cite{Lotshawdataset,lotshaw2021bfgs}. Data collected over various system sizes are utilized to extract an effective scaling of the average expectation value as a function of the noise parameter and optimal QAOA order, which is a function of system size. In the weak noise limit, the expectation value decays exponentially according to
\begin{align}
    \braket{H_C}&\approx \braket{H_C}_0 \exp\left[-\chi_{H_C}(p^*)\right]\\
    \textrm{where, }\chi_{H_C}(p^*)&\approx
    \begin{cases} 
    -\Gamma p^*/\Gamma_s, & \textrm{[stochastic error]} \\ 
    -\eta^2 p^*/\eta_c^2, & \textrm{[coherent error]} \end{cases}
\end{align}
with $\sqrt{\Gamma_s}=0.51$ and $ \eta_c=0.46$. Note that in the case of large stochastic noise, $\Gamma p^* \gg 1$, we see that the noise-averaged cost function approaches zero in a power-law, $\overline{\braket{H_C}}\propto 1/(\Gamma p^*)^\alpha$, with $\alpha=1.46$.

The scaling fit is compared against the cumulant expansion and its associated bounds in Fig.~\ref{fig:grover-bounds}(b). In general, this comparison will require a similar approach to that used in the Grover's search analysis. Namely, we must calculate the approximate dynamics and bounds by summing the contribution of each constituent $Z_iZ_{i+1}$ to the faulty expectation value of $H_C$. However, for the case considered here, $n=10$, the inverse of $H_C$ does exist. Thus, no summation is required in the calculation of $\overline{\braket{H_C}}$ and the bounding expression given in Eq.~(\ref{eq:abs-error-ev}) directly applies.

Overall, we find good agreement between the second cumulant dynamics and the fit in the weak noise regime. Numerically calculated bounds (blue triangles) and second cumulant dynamics (purple circles) track closely, indicating that the bound is tight for this particular problem class. The analytical bound (green dashed line) is determined by utilizing
\begin{equation}
    \|C^{(2)}_{H_C}(T)\|_{\infty}\leq n\Gamma \left(1+\frac{3n}{\sigma_{\min}}\right)\sum^{n/2}_{i=1} (|\gamma_i| + |\beta_i|),
\end{equation}
where $\sigma_{\min}$ is the minimum singular value of $H_C$, to further bound Eq.~(\ref{eq:abs-error-ev}). As expected, this additional bound allows for general behavior of the dynamics to be captured, but does not provide a tight error bound, most notably for increasing variance.





\section{Digitization: Avoiding Precision-induced errors in QAOA} \label{sec:digitization}
Let us now discuss some general strategies to mitigate precision-induced errors in QAOA. It is quite clear that precision errors in QAOA variational parameters (whether coherent or stochastic) can be quite detrimental to the performance of the algorithm. Constant coherent over- or under-rotations can be straightforwardly accounted for by updating the optimized angles exactly by the coherent error.  However, stochastic errors are harder to deal with  using this approach as they are random in nature.  Below, we propose a method for mitigating the effects of stochastic precision errors that involves digitizing the angles in a binary representation.

From the analysis of the previous section, it is clear that for the implementation of a $p$-layer QAOA algorithm to achieve a desired accuracy threshold for the expectation value of the cost function, $|\Delta\braket{H_C}|/\braket{H}_0=2^{-\epsilon}$ ($\epsilon>0$), the maximum precision error permitted in the variational parameters is $\eta_H,\eta_C \propto \sqrt{\epsilon/p}$, where the constant pre-factor depends on the details of the problem. Clearly, the precision requirements depend on the problem type. For implementing the Grover search algorithm, the optimal number of layers necessary scales exponentially with the number of qubits, and thus the required precision  must increase at least as $2^{-n/2}$. We expect this scenario to be fairly generic for QAOA circuits. While short finite-depth QAOA circuits are currently being investigated for NISQ applications in optimization, it is expected that for computationally hard optimization problems it is not unreasonable to expect the number of layer to increase at least super-polynomially with problem size~\cite{qnp}. 

 The digitization of the QAOA angles may allow us to avoid errors induced by precision errors (at the cost of increasing circuit depth). 
 Let us start by discussing how to implement the QAOA cost function and mixer unitaries up to a certain precision error in the variational parameters. Consider an $N_\gamma$ bit implementation of the angle $\gamma$ in the unitary $\exp(-i\gamma H_C)$,
\begin{align}
    \gamma\approx \frac{2\pi}{2^{N_\gamma}} \sum_{j=1}^{N_\gamma} A_j 2^{j}
\end{align}
where, $A_j\in \{0,1\}$ is the $j$-th bit of the integer closest to $\gamma\frac{2^N}{2\pi}$. Note that the $N_\gamma$ bit implementation has a precision error $\sim 2^{-N_\gamma}$. The unitary may be implemented as,
\begin{align}
    U_C(\gamma)=\exp\left(-i\gamma H_C\right)=\prod_{j=1}^{N_\gamma}\left[\exp\left(i\frac{2\pi}{2^N} 2^j H_C\right)\right]^{A_j} \label{eq:Uc_precision}
\end{align}
A similar analysis can also be extended to the mixer unitary, $U_M(\beta)$. Now, instead of viewing QAOA as an analog quantum algorithm with continuously variable angles, one only needs to implement  certain \emph{below-threshold} building block unitaries, $U_{Cj}=\exp\left(i\frac{2\pi}{2^N} 2^j H_C\right)$ and $U_{Mj}=\exp\left(i\frac{2\pi}{2^N} 2^j H_M\right)$, with $j=0,1\cdots,N_\gamma$ or $N_\beta$. (The term ``below-threshold" denotes unitaries that can be implemented with a precision error of at most $\sim \sqrt{\epsilon/p N_{\gamma,\beta}}$.) Combined with Eq.~(\ref{eq:Uc_precision}), both the unitary operators can be implemented within the desired precision at the cost of increasing the circuit depth $\sim p N_\gamma C_\gamma +p N_\beta C_\beta$, where  $C_{\gamma.\beta}$ are the maximum circuit depth necessary to realize one of the unitary operations $U_{Cj}$, $U_{Mj}$.

The implementation of a $p$-layer QAOA circuit with the desired cost-function evaluation accuracy of $2^{-\epsilon}$ requires a precision error of at most $\sim \sqrt{\epsilon/p}$. From the analysis above, an $N_{\gamma,\beta}$ bit implementation of the variational parameters introduces a precision error $\propto 2^{-N_{\gamma,\beta}}$ and thus, we conclude that the effects of precision errors larger than $\sqrt{\epsilon/p}$ may be mitigated by $N_{\gamma,\beta}\propto |\log_2(\sqrt{\epsilon/p})|$. As a result, mitigating precision errors requires an increase in the circuit depth that is logarithmic with the desired $\epsilon$ and $p$. 
This would imply that for problems that are hardest for QAOA, where $p\propto 2^n$ (such as Grover's Search), one can avoid precision-errors at a cost polynomial ($\propto n$) in the number of qubits. 

\section{Conclusions}
\label{sec:concl}
In this study, we investigated the effect of precision errors, or the misspecification of variational parameters, on QAOA performance. We provided insight into the extent of their harm on parameter training and performance guarantees. Utilizing concepts from quantum control theory, we analytically estimated the contribution of precision errors to QAOA dynamics. This approach enabled the development of bounds on expectation values, approximation ratios, and parameter training error. Through our analysis, we found that any fixed precision implementation of QAOA will realize performance guarantees that greatly differ from the idealized setting. In particular, we found an exponential reduction in success probability with increasing QAOA order and error magnitude. Numerical studies of the QAOA variant of Grover's search and the one-dimensional transverse-field Ising model conveyed these observations and provided verification for our analytical estimates. 

Despite the detrimental nature of precision errors on QAOA performance, we showed that it is possible to mitigate such errors by digitizing the variational parameters. Each constituent QAOA evolution operator was then expanded into a product of operators whose circuit depth is determined by the desired precision. Provided that the decomposition is expressed in terms of evolution operators that can be implemented with a greater precision than that required by the algorithm, one can successfully achieve a desired precision accuracy for QAOA.

While our study focused on precision errors, the analytical approach inherently extends to a far greater class of noise models. Specifically, one can utilize this approach to study the effect of spatial and temporally correlated noise on QAOA and variational algorithms more generally. In fact, the bounds we developed in this study naturally extend to these more general scenarios without further manipulation. In summary, we view this framework as providing a unique perspective on variational quantum algorithms and as a constructive tool for predicting faulty algorithm performance and facilitating the development of noise-robust variants.

\section{Acknowledgements}
Gregory Quiroz, Paraj Titum, Pavel Lougovski, Kevin Schultz, Eugene Dumitrescu, and Itay Hen acknowledge funding from the U.S. Department of Energy (DOE), Office of Science, Office of Advanced Scientific Computing Research (ASCR) Quantum Computing Application Teams program, under fieldwork proposal number ERKJ347.  Phillip Lotshaw was supported at ORNL by the Defense Advanced Research Project Agency, Defense Science Office under contract HR001120C0046 with Georgia Tech Research Institute.

\appendix
\section{Alternative Toggling Frame}
\label{sec:alt-toggling}
Faulty QAOA evolution dictated by Eq.~(\ref{eq:H-f-QAOA}) can be factorized according to Eq.(\ref{eq:u-rep}), or equivalently, $U(T)=U_0(T)\tilde{U}^\prime_E(T)$. The latter expression follows a very typical interaction picture representation in which the time evolution operator
\begin{equation}
    \tilde{U}^\prime_E(T) = \mathcal{T}_{+}e^{-i\int^T_0 \tilde{H}^\prime_E(t) \, dt}
\end{equation}
is governed by $\tilde{H}^\prime_E(t) = U^\dagger_0(t)H_E(t) U_0(t)$. Equivalence, between the two representations can be shown in a very straightforward manner by observing that
\begin{eqnarray}
U(T)&=&U_0(T)\tilde{U}^\prime_E(T)\nonumber\\
&=& U_0(T)\tilde{U}^\prime_E(T)U^\dagger_0(T)U_0(T)\nonumber\\
&=&\tilde{U}_E(T) U_0(T);
\end{eqnarray}
thus, matching the expressions in Eq.~(\ref{eq:u-tot})-(\ref{eq:u-err}).

Both representations lend themselves to cumulant-based expansions of expectation values with the latter inducing a time-dependence on the observable. Following the approach utilized in the main text, the expectation value of an observable $O$ can be expressed in terms of $U(T)=U_0(T)\tilde{U}^\prime_E(T)$ as
\begin{eqnarray}
    \braket{O} &=& \overline{\Tr{[O\rho(T)]}} \nonumber\\
    &=&\overline{\Tr{[O U_0(t)U_E(t)\rho(0)(\tilde{U}^\prime_E(t))^\dagger U^\dagger_0(t)]}}\nonumber\\
    &=&\Tr{[\Lambda(t) \rho(0)\tilde{O}(t)]}.
    \label{eq:alt-exp}
\end{eqnarray}
The error dynamics are determined by the operator $\Lambda(t)=\overline{\tilde{O}^{-1}(t)U^\dagger_E(t)\tilde{O}(t)U_E(t)}$, which includes a conjugation by $\tilde{O}(t)=U^\dagger_0(t) O U_0(t)$, the observable in the interaction picture with respect to the noiseless QAOA evolution. As can be seen from comparing the above expressions to those surrounding Eq.~(\ref{eq:avg_O_lamb}), the distinction lies in the definition of the observable. Note that while both representations are analogous, the additional time-dependence can result in unnecessary complexity in analytical or numerical investigations. Hence, we select the former for this study.


\section{Additional Bounds}
\subsection{Absolute Error Bound: Non-Invertible Observable}
\label{subsec:noninv-o}
Consider the case where an observable $O$ is non-invertible, however, it can be expressed as a sum of invertible operators $O_i$: $O=\sum_i O_i$. A trivial example of an operator expansion of this type is the $n$-qubit Pauli basis for an operator $O\in SU(n)$. Of course, this particular example is relevant to the Grover problem and Ising-type problem considered in the main text and, more generally, in the context of variational quantum algorithms.

Utilizing this expansion for $O$, the expectation value of the observable can be written as
\begin{eqnarray}
\overline{\braket{O(T)}} &=& \sum_i \overline{\Tr{[\rho(T) O_i]}}\nonumber\\
&=& \sum_i \Tr{[\Lambda_i(T)\rho_S(T) O_i]}, 
\end{eqnarray}
where $\Lambda_i(T)=\overline{O^{-1}_i \tilde{U}^\dagger_E(T) O_i \tilde{U}_E(T)}$. Going a bit further and leveraging the cumulant expansion,
\begin{equation}
    \overline{\braket{O(T)}} = \sum_i \tr{e^{\mathcal{C}_{O_i}(T)}\rho_S(T) O_i},
\end{equation}
with $\mathcal{C}_{O_i}(T)$ representing the cumulant expansion containing the $j$th observable.

The above expressions can be employed to derive bounds on the various quantities examined in Sec.~\ref{sec:bounds} of the main text. For example, consider the absolute error in the expectation value $|\overline{\Delta\braket{O}}|$. An upper bound on this quantity can be derived following a similar approach to that of Sec.~\ref{subsec:err-ev} to yield
\begin{eqnarray}
|\overline{\Delta\braket{O(T)}}| &=& \left|\tr{\left(\sum_i O_i\Lambda_i(T)-O\right)\rho_S(T)}\right| \nonumber\\
&\stackrel{(1)}{\leq}& \| \sum_i O_i\Lambda_i(T)-O \|_\infty \,\,\|\rho_S(T)\|_1\nonumber\\
&\stackrel{(2)}{\leq}& \sum_i \|O_i\|_\infty \|\Lambda_i(T)-I\|_\infty\nonumber\\
&\stackrel{(3)}{\leq}& \sum_i \|O_i\|_\infty \left(e^{\|\mathcal{C}_{O_i}(T)\|_\infty}-1\right).
\end{eqnarray}
Here, the variant of H\"older's inequality from Sec.~\ref{subsec:err-ev} is used to obtain (1). The trace-norm of a density matrix is unity and thus, the resulting bound in (2) is solely dependent upon the infinity norm of a term proportional to the accumulated error resulting from $\Lambda_i(T)$. In addition to the property of density matrix, we apply the triangle inequality and sub-multiplicativity to obtain (2). By expressing the error operator as a cumulant expansion and bounding the error term as in Sec.~\ref{subsec:bd-lambda}, we obtain the final bound in (3).

\subsection{Bounds on Cumulants}
\label{subsec:cum-bounds}
Many of the bounds derived above share a common feature: dependence on the operator norm of the error operator. Through Eq.~(\ref{eq:L-bound}), we find that the error dynamics can be bounded in terms of the operator norm of the cumulant series $\mathcal{C}_O(T)$. When the noise is sufficiently weak, the cumulant series is well-approximated by the leading order cumulant expression, i.e., $\mathcal{C}^{(1)}_O(T)$ or $\mathcal{C}^{(2)}_O(T)$ in the case of constant coherent errors or zero-mean, Gaussian stochastic errors, respectively. Here, we provide bounds on the first and second cumulant which can be used to further upper bound the expressions presented above. 

Using the expressions derived in Sec.~\ref{sec:err-an} for the first and second cumulants, bounds are obtained for each noise scenario. In the case of constant coherent errors, the first cumulant [Eq.~(\ref{eq:c1})] can be shown to be bounded according to
\begin{equation}
    \left\|\mathcal{C}^{(1)}_O\right\| 
    \leq \sum_{\mu=M,C}\eta_\mu \sum^{p}_{j=1}  |g^\mu_j| \|H_\mu\| \left( 1  + \|O^{-1}\| \|O\|\right).
    \label{eq:c1-bound}
\end{equation}
Utilizing similar techniques, such as the triangle inequality, sub-multiplicativity, and unitary invariance, a similar expression can be obtained for the second cumulant [Eq.~(\ref{eq:c2})]:
\begin{eqnarray}
    \left\|\mathcal{C}^{(2)}_O\right\| 
    &\leq&\|I_1(T)\| + \|I_2(T)\| + 2\|I_3(T)\|\nonumber\\
    &\leq& \sum_{\mu=M,C} \Gamma_\mu \sum^{p}_{j=1}  |g^\mu_j| \|H_\mu\| \left(1 + 3\|O^{-1}\|\|O\|\right).\quad\quad
    \label{eq:c2-bound}
\end{eqnarray}
For brevity and without loss of generality, the subscript denoting the operator norm has been dropped and bounds are displayed in terms of a generic unitary invariant norm.

Intuitively, bounds on the first and second cumulant scale with mean and variance, respectively. Both bounds are further characterized by the accumulative time over which each QAOA constituent Hamiltonian contributes to the evolution. Additional dependence on the norm of the of the constituent Hamiltonians and the observable $O$ and its inverse are also observed. The latter emerging via integral expressions conjugated by $O$.

\subsection{Bounds on Gradients of Cumulants}
Here, we include bounding expressions for $\|\partial_{\gamma_k}\mathcal{C}^{(n)}_O(T)\|$ given their relevance to the gradient bound derived in Eq.~(\ref{eq:lambda}). While we focus of the derivative with respect to $\gamma_k$, we note that equivalent expressions for the derivative with respect to $\beta_k$ follow directly. The primary difference being a change of QAOA Hamiltonian from $H_C$ to $H_M$ at the appropriate locations.

In order to bound the derivatives of the cumulants, we begin by noting that 
\begin{equation}
    \partial_{\gamma_k} Q_{p:j+1} = \left\{
    \begin{array}{lcr}
    -i Q_{p:k}H_C Q_{k-1:j+1} &:& k\geq j+1\\
    0 &:& \text{otherwise}
    \end{array}\right..
    \label{eq:partial-q}
\end{equation}
This expression can be utilized to bound on the derivative of the rotated-frame Hamiltonians as
\begin{eqnarray}
    \|\partial_{\gamma_k}\left(Q_{p:j+1}H_C Q^\dagger_{p:j+1}\right)\| 
    &\leq& 2 \|H_C\|^2,
\end{eqnarray}
for $k\geq j+1$, which in turn can be employed to bound both cumulants. The first cumulant is bounded according to
\begin{eqnarray}
    \|\partial_{\gamma_k} \mathcal{C}^{(1)}_O(T)\| &\leq& \left(1+\|O^{-1}\|\|O\|\right) \|H_C\|\nonumber\\
    & &\times \left[ 
     2\eta_C\|H_M\|\sum^{k-1}_{j=1} |\beta_j|\right. \nonumber\\
    & & \left. +\eta_M \left(1+2\|H_C\|\sum^{k}_{j=1} |\gamma_j|\right)\right],\quad
\end{eqnarray}
where the sums over $j\leq k-1$ arises from Eq.~(\ref{eq:partial-q}). The derivative of the second cumulant is bounded as
\begin{eqnarray}
    \|\partial_{\gamma_k}\mathcal{C}^{(2)}_O(T)\| &\leq& \|\partial_{\gamma_k}I_1(T)\| + \|\partial_{\gamma_k}I_2(T)\|+ 2\|\partial_{\gamma_k}I_3(T)\| \nonumber\\
    &\leq& \|\partial_{\gamma_k}I_1(T)\| (1+\|O^{-1}\| \|O\|) \nonumber\\
    & & + 2\|\partial_{\gamma_k}I_3(T)\|,
\end{eqnarray}
in terms of its components integrals. In turn, these integrals can be further bounded in terms of the variational parameters and the norm of the mixer and problem Hamiltonian. The first integral bound is
\begin{eqnarray}
    \|\partial_{\gamma_k}I_1(T)\|&\leq& \Gamma_M \|H_C\|^2 \left( 1+2\|H_C\| \sum^k_{j=1}|\gamma_j|\right)\nonumber\\
    & & + 2\Gamma_C\|H_M\|^2\|H_C\| \sum^{k-1}_{j=1}|\beta_j|,
\end{eqnarray}
while the derivative of the third integral is bounded by
\begin{eqnarray}
    \|\partial_{\gamma_k}I_3(T)\| &\leq& \|O^{-1}\| \|O\|\|H_C\|\nonumber\\
    & &\times \left[ \Gamma_C\|H_C\| \left(1 + 4\|H_C\|\sum^k_{j=1}|\gamma_j|\right)\right. \nonumber\\
    & & \left.+ 4\Gamma_M \|H_M\|^2\sum^{k-1}_{j=1}|\beta_j|\right].
\end{eqnarray}

\subsection{Finite Sampling Effects}
Finite sampling of expectation values will result in estimates that deviate from the asymptotic, infinite sample mean. As such, we account for finite sampling effects in the bounds derived in Sec.~\ref{sec:bounds} by modifying the estimator in accordance with the Central Limit Theorem to
\begin{equation}
    \braket{O}_{\rm est} = \braket{O} \pm \frac{1}{\sqrt{N_s}}\sqrt{{\rm Var}(O)}.
    \label{eq:ev-sampling}
\end{equation}
By definition, this assumes that the sampling statistics of the expectation value are Gaussian. We make use of Eq.~(\ref{eq:ev-sampling}) below to derive finite sampling corrections to the bounds on absolute error and mean-squared error.

\subsubsection{Absolute Error}
Here, we attain a bound on the absolute error between the finitely sampled noisy expectation value $\braket{O}_{\rm est}$ and the asymptotic noiseless expectation value $\braket{O}_0$. Following Eq.~(\ref{eq:abs-err}), this quantity is defined as
\begin{equation}
    |\overline{\Delta\braket{O(T)}}_{\rm est}| = |\overline{\braket{O(T)}}_{\rm est}-\braket{O(T)}_0|.
\end{equation}
Using Eq.~(\ref{eq:ev-sampling}), this expression can be bounded by the sum of the asymptotic bound for $|\overline{\Delta \braket{O(T)}}|$ [Eq.~(\ref{eq:abs-error-ev})] and a correction term proportional to $1/\sqrt{N_s}$. More concretely, it can be shown that
\begin{eqnarray}
    |\overline{\Delta\braket{O(T)}}_{\rm est}| - |\overline{\Delta\braket{O(T)}}| &\leq& \frac{1}{\sqrt{N_s}}|\sqrt{{\rm Var}(O)}|\nonumber\\
    &\leq& \frac{1}{\sqrt{N_s}}\sqrt{\kappa(O,T)},
    \label{eq:abs-err-fs}
\end{eqnarray}
where
\begin{equation}
    \kappa(O,T) = \|\Lambda(T)\|_\infty \|O^2\rho_0(T)\|_1 + \|\Lambda(T)\|^2_\infty \|O\rho_0(T)\|^2_1
\end{equation}
results from upper bounding the variance of $O$. 

\subsubsection{MSE}
Following Eq.~(\ref{eq:mse}), the MSE in the finite sampling regime is defined as
\begin{equation}
    \overline{{\rm MSE}(O(T))}_{\rm est} = \overline{{\rm Var}_{\rm est}(O(T))} + \left(\overline{\Delta\braket{O(T)}}_{\rm est}\right)^2,
\end{equation}
where the estimator is now designated as the finitely sampled expectation value given in Eq.~(\ref{eq:ev-sampling}). Similar to the bound on the bias developed in Eq.~(\ref{eq:abs-err-fs}), the variance can be shown to be bounded by the sum of the asymptotic variance and $N_s$-dependent corrections, i.e.,
\begin{eqnarray}
    {\rm Var}_{\rm est}(O) &=& \braket{O^2}_{\rm est} - \left(\braket{O}_{\rm est}\right)^2\nonumber\\
    &\leq& {\rm Var}(O) + \frac{1}{N_s}{\rm Var}(O)\nonumber\\
    & & +\frac{1}{\sqrt{N_s}}[\sqrt{{\rm Var}(O^2)} + \braket{O}\sqrt{{\rm Var}(O)}]. \quad\quad
\end{eqnarray}
The relative difference between the finitely sampled and asymptotic variance can be further bounded according to
\begin{eqnarray}
    \overline{{\rm Var_{\rm est}}(O)} -  \overline{{\rm Var}(O)}&\leq& \frac{1}{N_s} \kappa(O,T)+ \frac{1}{\sqrt{N_s}}\left[\kappa(O^2,T)\right. \nonumber\\
    & & \left. + 2\|\Lambda(T)\|_\infty \|O\rho_0(T)\|_1 |\sqrt{\kappa(O,T)}\right].\nonumber\\
\end{eqnarray}
As expected by the definition of Eq.~(\ref{eq:ev-sampling}), the dominant term scales according the $1/\sqrt{N_s}$.

\section{Derivations}
\label{app:derivations}
Here, we provide additional detailed regarding the bounds presented in Sec.~\ref{sec:bounds}.

\subsection{Bound on Error Operator}
\label{subsec:bd-lambda}
In Sec.~\ref{subsec:err-ev}, the operator norm of the operator $\Lambda(T)$ was said to be bounded by the exponentiated norm of the cumulant series. We prove this bound as follows:
\begin{eqnarray}
\|\Lambda(T)\| &=& \|e^{\mathcal{C}_O(T)}\|\nonumber\\
&=& \|\sum^\infty_{m=0}\mathcal{C}^m_O(T)/m!\|\nonumber\\
&\stackrel{(1)}{\leq}& \sum^\infty_{m=0}\|\mathcal{C}^m_O(T)/m!\|\nonumber\\
&\stackrel{(2)}{\leq}& \sum^\infty_{m=0}\|\mathcal{C}_O(T)\|^m/m!\nonumber\\
&=&e^{\|\mathcal{C}_O(T)\|}.
\end{eqnarray}
Here, (1) triangle inequality and (2) sub-multiplicativity have been used to obtain the upper-bounding expressions. 

\subsection{Bound on MSE: Coherent Error}
MSE can be expressed as a sum of the variance and the bias of the estimator. Here, we derive bounds on each term individually for the case of constant coherent errors. The resulting bounds are combined to obtain Eq.~(\ref{eq:mse-bound-const}). Although the bound derived in the subsequent subsection encompasses constant coherent errors, we take an alternative approach that leverages results from Trotter error analysis to obtain a tighter bound for a case where ensemble averaging (i.e., a cumulant expansion) is not required.

A bound on the variance can be obtained by first noting that
\begin{eqnarray}
{\rm Var}(O(T)) &=& \Tr{[\rho(T) O^2]} - \Tr{[\rho(T) O]}^2 \nonumber\\
&\leq& |\Tr{[\rho(T) O^2]}| + |\Tr{[\rho(T) O]}|^2,
\end{eqnarray}
where we have used ${\rm Var}(O)=|{\rm Var}(O)|$ and the triangle inequality. Employing H\"older's inequality and sub-multiplicativity, this bound is further reduced to
\begin{eqnarray}
    {\rm Var}(O(T)) &\leq& \|O^2\|_\infty + \|O^2\|^2_\infty \nonumber\\
    &\leq& 2\|O\|^2_\infty.
    \label{eq:app-var-const}
\end{eqnarray}
Note that the first inequality makes use of $\|\rho(T)\|_1=1$.

The squared-bias is bounded by first bounding the difference between the faulty and noiseless expectation values. Following a similar procedure to that utilized above,
\begin{eqnarray}
|\Delta \braket{O(T)}| &=& |\Tr{[\rho(T)-\rho_0(T)O]}|\nonumber\\
&=& |\Tr{[(U^\dagger(T)OU(T) - U^\dagger_0(T)OU_0(T)) \rho(0)]}| \nonumber\\
&\stackrel{(1)}{\leq}& \|(U^\dagger(T)OU(T) - U^\dagger_0(T)OU_0(T))\|_\infty.
\end{eqnarray}
We now introduce an additional term, $U^\dagger(T)OU_0(T) - U^\dagger(T)OU_0(T)$, within the operator norm and group terms associated with the difference $U^\dagger(T) - U_0(T)$ and its Hermitian conjugate. Using unitary invariance of the norm and sub-multiplicativity, we obtain
\begin{equation}
|\Delta \braket{O(T)}| \leq 2\|O\|_\infty \|U(T)-U_0(T)\|_\infty.
\label{eq:app-bias-bound-const}
\end{equation}

The norm of the difference between the faulty and noiseless propagators is further bounded through the definition of the time evolution dynamical equation. Namely,
\begin{eqnarray}
    \|U(T)-U_0(T)\|_\infty &=& \|\tilde{U}_E(T) - 1\|_\infty \nonumber\\
    &=& \|-i \int^T_0 \tilde{H}_E(t) U_E(t)\,\,dt\|_\infty \nonumber\\
    &\leq& \int^T_0 \|H_E(t)\|_\infty \,\,dt \nonumber\\
    &\leq& h_E,
    \label{eq:app-u-diff}
\end{eqnarray}
where the triangle inequality and sub-multiplicativity are used to achieve the final bound in terms Eq.~(\ref{eq:he}). Including Eq.~(\ref{eq:app-u-diff}) into the Eq.~(\ref{eq:app-bias-bound-const}) leads to $(\Delta \braket{O(T)})^2\leq 4 \|O\|^2_\infty h^2_E$, which in turn yields Eq.~(\ref{eq:mse-bound-const}) when combined with Eq.~(\ref{eq:app-var-const}).

\subsection{Bound on Exact Gradient}
In Sec.~\ref{subsec:err-training}, we briefly derived a bound on the absolute error in the gradient of the expectation value for a single variational parameter. We elaborate on that derivation here, focusing specifically on the expressions related to norms comprised of the derivative of the error operator and the derivative of the ideal QAOA time-evolved state. The former appears in the bound derived in Eq.~(\ref{eq:lambda}), while the latter is a part of Eq.~(\ref{eq:grad-2}).

A bound on the norm of the derivative of the error operator is shown in Eq.~(\ref{eq:lambda}). We elucidate the details on the derivation of this bound as follows:
\begin{eqnarray}
\|\partial_{\gamma_k}\Lambda(T)\| &=& \left\|e^{\mathcal{C}_O(T)}\int^{1}_{0}d\lambda\,\, e^{-\lambda \mathcal{C}_O(T)}\partial_{\gamma_k}\mathcal{C}_O(T)e^{\lambda \mathcal{C}_O(T)}\right\| \nonumber\\
&\leq&\|e^{\mathcal{C}_O(T)}\| \int^{1}_0 d\lambda \left\| e^{-\lambda \mathcal{C}_O(T)}\partial_{\gamma_k}\mathcal{C}_O(T)e^{\lambda \mathcal{C}_O(T)}\right\| \nonumber\\
&\leq& e^{\|\mathcal{C}_O(T)\|} \int^{1}_0 e^{2\lambda \|\mathcal{C}_O(T)\|}d\lambda \,\, \|\partial_{\gamma_j}\mathcal{C}_O(T)\|\nonumber\\
&=& \frac{e^{\|\mathcal{C}_O(T)\|}}{2\|\mathcal{C}_O(T)\|}\left(e^{2\|\mathcal{C}_O(T)\|}-1\right)\|\partial_{\gamma_k}\mathcal{C}_O(T)\|.
\end{eqnarray}
The first equality is obtained by expressing the error operator as a cumulant expansion, i.e., $\Lambda(T)=e^{\mathcal{C}_O(T)}$ and then using an identity derived by Snider~\cite{snider1964:pert} to formally calculate the derivative of an exponential operator. Subsequent inequalities follow from applications of sub-multiplicativity and Eq.~(\ref{eq:L-bound}), while the final equality results from an evaluation of the integral.

While the first term in the bound given in Eq.~(\ref{eq:grad-partial}) is proportional to the derivative of the error operator, the second is proportional to $\partial_{\gamma_k}\rho_0(T)$. The subsequent bound on this term [Eq.~(\ref{eq:grad-2})] is first computed by noting that 
\begin{eqnarray}
\partial_{\gamma_k}\rho_0(T) &=& (\partial_{\gamma_k}U)\rho(0)U^\dagger + U\rho(0)(\partial_{\gamma_k}U^\dagger) \nonumber\\
&=& -i Q_{p:k+1}\left[H_C,  Q_{k:1} \rho(0) Q^\dagger_{k:1} \right]Q^\dagger_{p:k+1} \nonumber\\
&=& -i Q_{p:k+1}\left[H_C,  \rho(T_k)  \right]Q^\dagger_{p:k+1}.
\label{eq:deriv-rho}
\end{eqnarray}
Using H\"older's inequality to bound the second term of Eq.~(\ref{eq:grad-partial}), we obtain an expression that contains a norm of the product of $O$ and $\partial_{\gamma_k}\rho_0(T)$. Employing the result from Eq.~(\ref{eq:deriv-rho}), one can show
\begin{eqnarray}
\|O \partial_{\gamma_k} \rho_0(T)\| &=& \|O Q_{p:k+1}\left[H_C,  \rho(T_k)  \right]Q^\dagger_{p:k+1}\| \nonumber\\
&=&\|Q^\dagger_{p:k+1} O Q_{p:k+1}\left[H_C,  \rho(T_k)  \right]\| \nonumber\\
&=&\|\tilde{O}(T,T_k)\left[H_C,  \rho(T_k)  \right]\|.
\end{eqnarray}
Note that the second equality follows from the unitary invariance of the norm.

\section{Additional Results for Numerical simulations.}
\label{app:numerics}
In this section, we provide additional details regarding the numerical simulations discussed in Sec.~\ref{sec:examples}. In Sec.~\ref{app:subsec:grover} we discuss additional results for the Grover problem and in in Sec.~\ref{app:subsec:ising} we provide additional information on the Ising problem. 
\subsection{Grover Problem} \label{app:subsec:grover}
\begin{figure*}
    \centering
    \includegraphics[width=\linewidth]{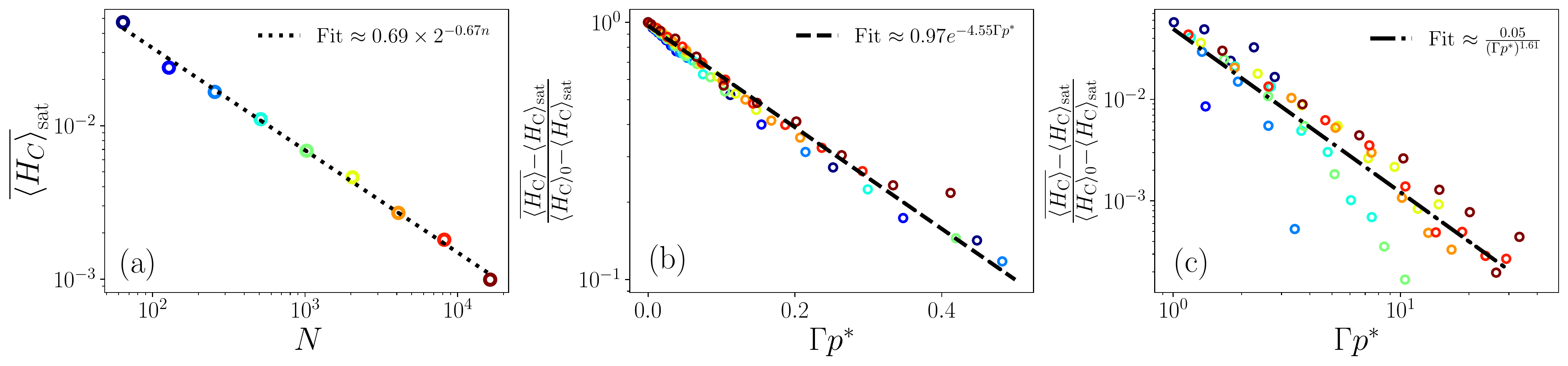}
    \caption{Scaling fits for the expectation value of the Grover cost Hamiltonian in the presence of stochastic errors [data corresponding to Fig.~\ref{fig:grover-size-dependence}(a)]. (a) shows the  fit to the saturation value for large noise. The saturation value decreases linearly with problem size, or exponentially with the number of qubits. (b) Weak noise ($\Gamma p^*<0.5$) behavior of the decay function (See Eq.~\ref{seq:decayfunc}). For weak noise, the expectation value decays exponentially as evident from the linear fit on the log-linear plot. (c) Large noise ($\Gamma p^*>1$) behavior of the decay function which behaves as a power-law. }
    \label{fig:averagedOverNormalErrorsGrover}
\end{figure*}
\begin{figure}
    \centering
    \includegraphics[width=\linewidth]{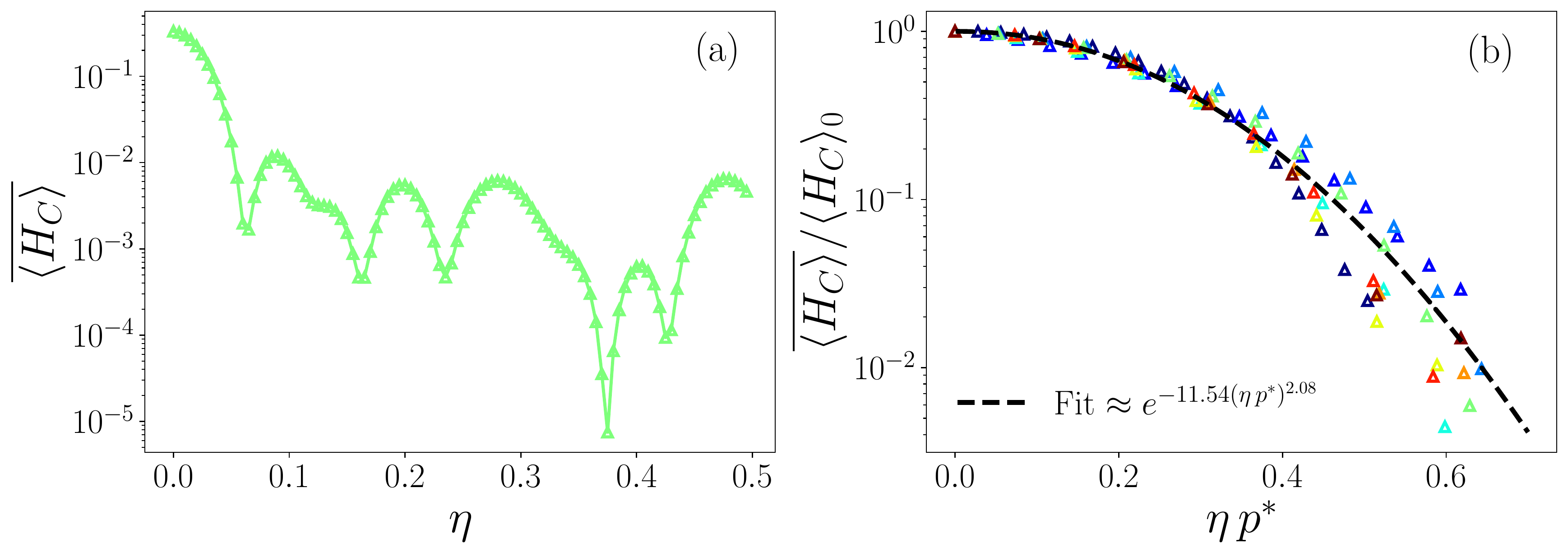}
    \caption{Scaling fits for the expectation value of the Grover cost function in the presence of coherent error [data corresponding to Fig.~\ref{fig:grover-size-dependence}(b)]. (a) shows the non-monotonic behavior of the noisy expectation value for increasing magnitude of the coherent error. (b) The decay of the expectation value for small coherent error $\eta p^*\lesssim 1$ behaves approximately as a squared exponential decay.}
    \label{fig:averagedOvercoherentErrorsGrover}
\end{figure}
\begin{figure}
    \centering
    \includegraphics[width=\linewidth]{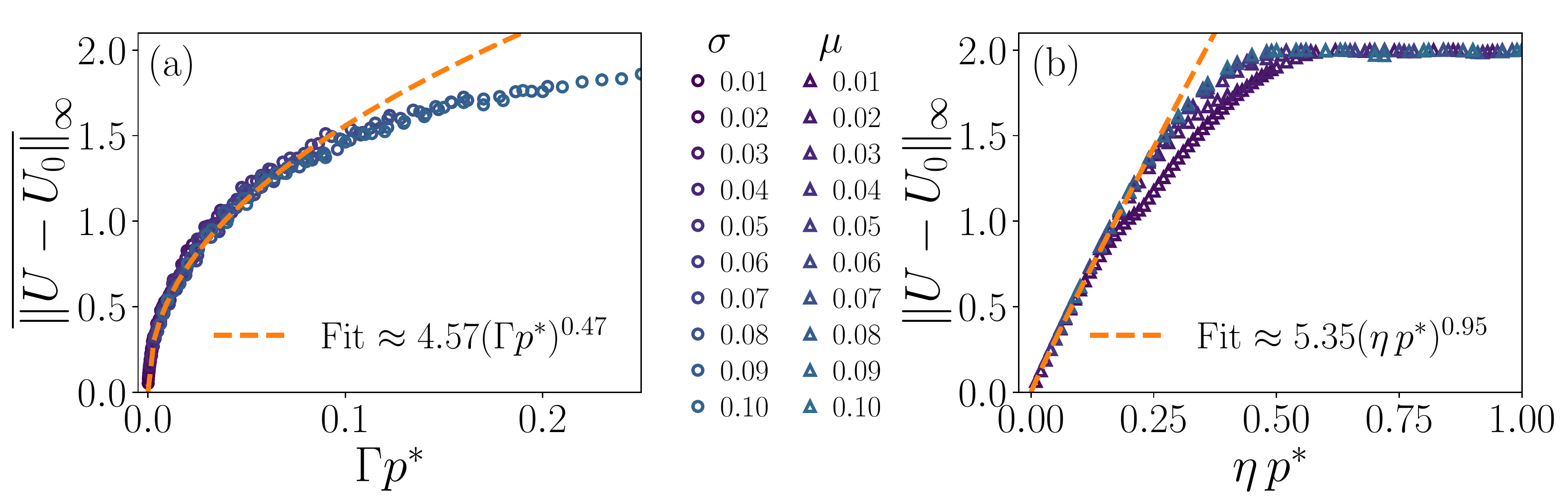}
    \caption{Scaling fits for the norm difference with increasing number of layers and noise, data corresponding to Figs.~\ref{fig:grover1} (c) and (d). (a) Diffusive growth in the distance for stochastic precision error. (b) Linear growth in the distance for coherent precision error. }
    \label{fig:normdifference}
\end{figure}

In this section we provide additional details for the fitting procedure used to extract the exponents discussed in Sec.~\ref{subsec:Grover}. We discuss the behavior of the noise-averaged expectation value in the presence of stochastic errors and coherent errors in Figs.~\ref{fig:averagedOverNormalErrorsGrover} and \ref{fig:averagedOvercoherentErrorsGrover} respectively. Finally, in Fig.~\ref{fig:normdifference} we discuss the fits to the distance between noisy and noiseless QAOA evolution.

Let us start by discussing the case of stochastic precision errors for the optimal Grover QAOA algorithm, the data for which is in Fig.~\ref{fig:grover-size-dependence}(a) in the main text. The results of the fitting procedure are discussed in Fig.~\ref{fig:averagedOverNormalErrorsIsing}. We fit the numerically evaluated noisy expectation value to the following phenomenological functional form that models both weak and strong noise behavior,
\begin{align}
    \overline{\braket{H_C}}=\braket{H_C}_0 \mathcal{S}_{H}(p^*) + \braket{H_C}_{\rm sat}\left[1-\mathcal{S}_{H}(p^*)\right] 
\end{align}
This functional form allows us to extract the decay function from the expression
\begin{align}
    \mathcal{S}_{H}(p^*)=\frac{ \overline{\braket{H_C}}-\braket{H_C}_{\rm sat}}{\braket{H_C}_0-\braket{H_C}_{\rm sat}},\label{seq:decayfunc}
\end{align}
with the condition, $\mathcal{S}_{H}(p^*=0)=1$. In Fig~\ref{fig:averagedOverNormalErrorsGrover}(a), we extract the functional form of the saturation value by examining the large-noise behavior ($\sigma=1$). We find that the saturation value decays exponentially with the number of qubits $n$ (or linearly with problem size $N$),
\begin{align}
    \overline{\braket{H_C}}_{\rm sat}\approx 0.69 \times 2^{-0.67 n}
\end{align}
Utilizing this expression for the large-noise saturation value, we obtain a fit for the decay function, $\mathcal{S}_H(\Gamma,p^*)$ in Fig.~\ref{fig:averagedOverNormalErrorsGrover}(b) and (c). We find two distinct functional behaviors for the weak noise [(b)] and large noise [(c)] respectively,
\begin{align}
    \mathcal{S}_{H}(p^*)=\begin{cases}
    0.97 e^{-4.55 \Gamma p^*},& \Gamma p^*<0.5  \textrm{ [weak noise]}\\
    \frac{0.05}{(\Gamma p^*)^{1.61}}, &\Gamma p^*>1.0  \textrm { [strong noise]}
    \end{cases}
\end{align}
Note that the weak and strong noise behavior is closely related to the behavior of the scaling variable $\sigma^2 p_{\rm opt}$.

Next, we extract the scaling behavior of the noisy expectation value in the presence of coherent errors, the data for which is shown in Fig.~\ref{fig:grover-size-dependence}(b) in the main text. The scaling fit is discussed in Fig.~\ref{fig:averagedOvercoherentErrorsGrover}. We note that measured expectation value in the presence of constant coherent error does not show a monotonic behavior with increase errors. This is shown in Fig.~\ref{fig:averagedOvercoherentErrorsGrover}(a) where we see that after an initial decay in the expectation value, the expectation value exhibits oscillations. We extract the initial exponential decay by examining the cases where $\eta p^*\lesssim 1$. The initial decay is well described by an exponential decay,
\begin{align}
\overline{\braket{H_C}}&=\braket{H_C}_0 {S}_{H}(p^*)\\
    \mathcal{S}_{H}(p^*)&= e^{-11.54(\eta\, p^*)^{2.08}}
\end{align}
Note that the expectation value decays as a squared-exponential.

Finally, let us discuss the scaling of the difference between the noisy and noiseless QAOA evolution operator. We numerically evaluate the $\infty$-norm of the average difference between the noisy unitary $U$ and the noiseless one, $U_0$ in Fig.~\ref{fig:normdifference}, the data for which corresponds to Fig.~\ref{fig:grover1} (c) and (d). We have the following behavior for weak noise or small $p$ ($p\sigma^2 \ll 1$,$\mu p \ll 1$),
\begin{align}
    \overline{\|U-U_0\|}_\infty &\approx
    \begin{cases}
      4.57(\Gamma p)^{0.47},& \textrm{[stochastic error]} \\ 
     5.35(\eta p)^{0.95}, &\textrm{[coherent error]} \\ 
    \end{cases}
\end{align}
This scaling is consistent with the norm-difference increasing in a diffusive manner for stochastic error ($\propto \sqrt{\Gamma p^*}$) and linearly for coherent error ($\propto \eta p^*$). 

\subsection{Ising Problem}
\label{app:subsec:ising}
\begin{figure}
    \centering
    \includegraphics[width=\linewidth]{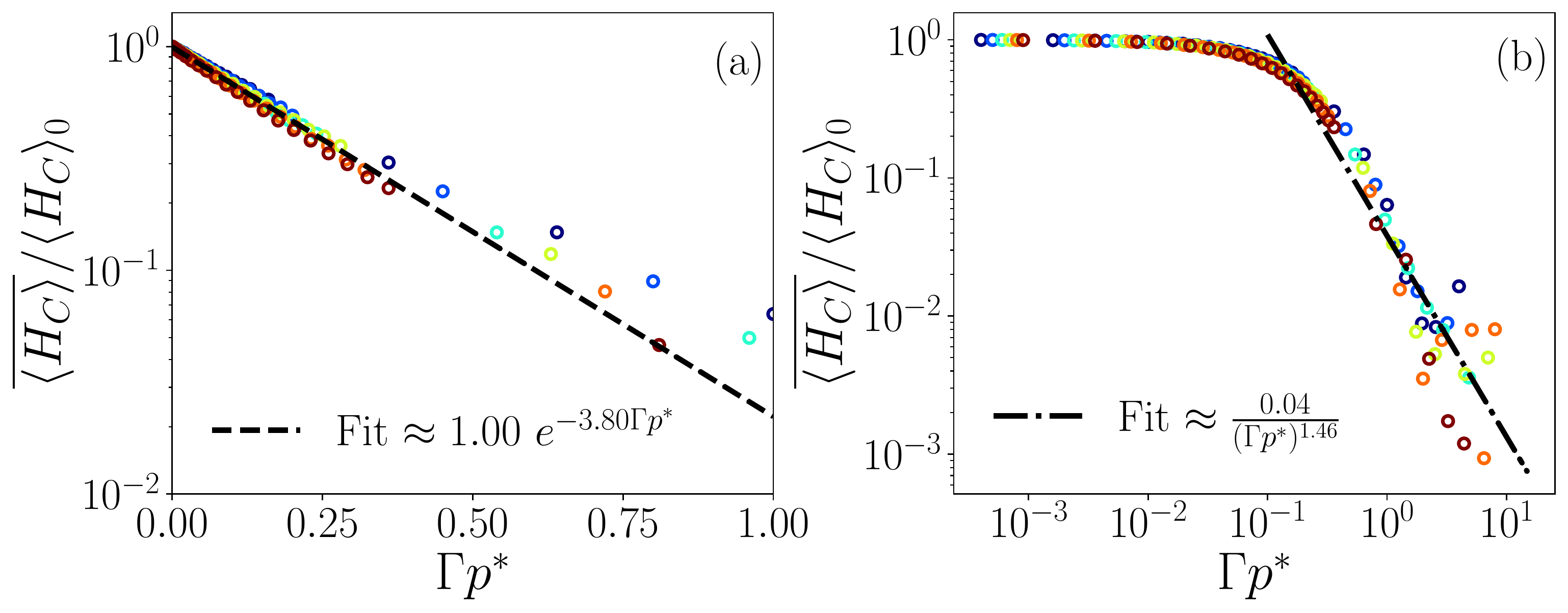}
    \caption{Scaling fits for GHZ Hamiltonian expectation value for stochastic errors.  (a) For weak noise ($\Gamma p^*\lesssim 1$) the expectation value of the Hamiltonian decays exponentially. (b) For strong noise ($\Gamma p^*\gtrsim 1 $), the expectation value decays as a power law.}
    \label{fig:averagedOverNormalErrorsIsing}
\end{figure}

In this section, we elaborate on the scaling fits for the Ising problem on a ring, that is discussed in Sec.~\ref{subsec:IsingInstances}. The data corresponds to Fig.~\ref{fig:ghz_size_dependence} in the main text. We fit the decay in the expectation value to the following phenomenological function,
\begin{align}
    \braket{H_C}&\approx \braket{H_C}_0  \mathcal{S}_H (p^*)
\end{align}
For stochastic error, as discussed in Fig.~\ref{fig:averagedOverNormalErrorsIsing}, the Hamiltonian expectation value decays as,
\begin{align}
    \mathcal{S}_{H}(p^*)=\begin{cases}
    1.00 e^{-3.80 \Gamma p^*},& \Gamma p^*<1.0  \textrm{ [weak noise]}\\
    \frac{0.04}{(\Gamma p^*)^{1.46}}, &\Gamma p^*>1.0  \textrm { [strong noise]}
    \end{cases}
\end{align}
Again, for stochastic errors, the expectation value decays exponentially for weak strength of the noise then changing to a power-law for larger strength of noise.

Next we discuss the effect of coherent errors. In this case, the decay function for weak noise is again consistent with an exponential decay,
\begin{align}
    \mathcal{S}_{H}(p^*)=e^{-4.64 \eta^2 p^*}.
\end{align}
Note that the exponential decay is consistent with a scaling variable $\eta^2 p^*$.
\begin{figure}
    \centering
    \includegraphics[width=0.8\linewidth]{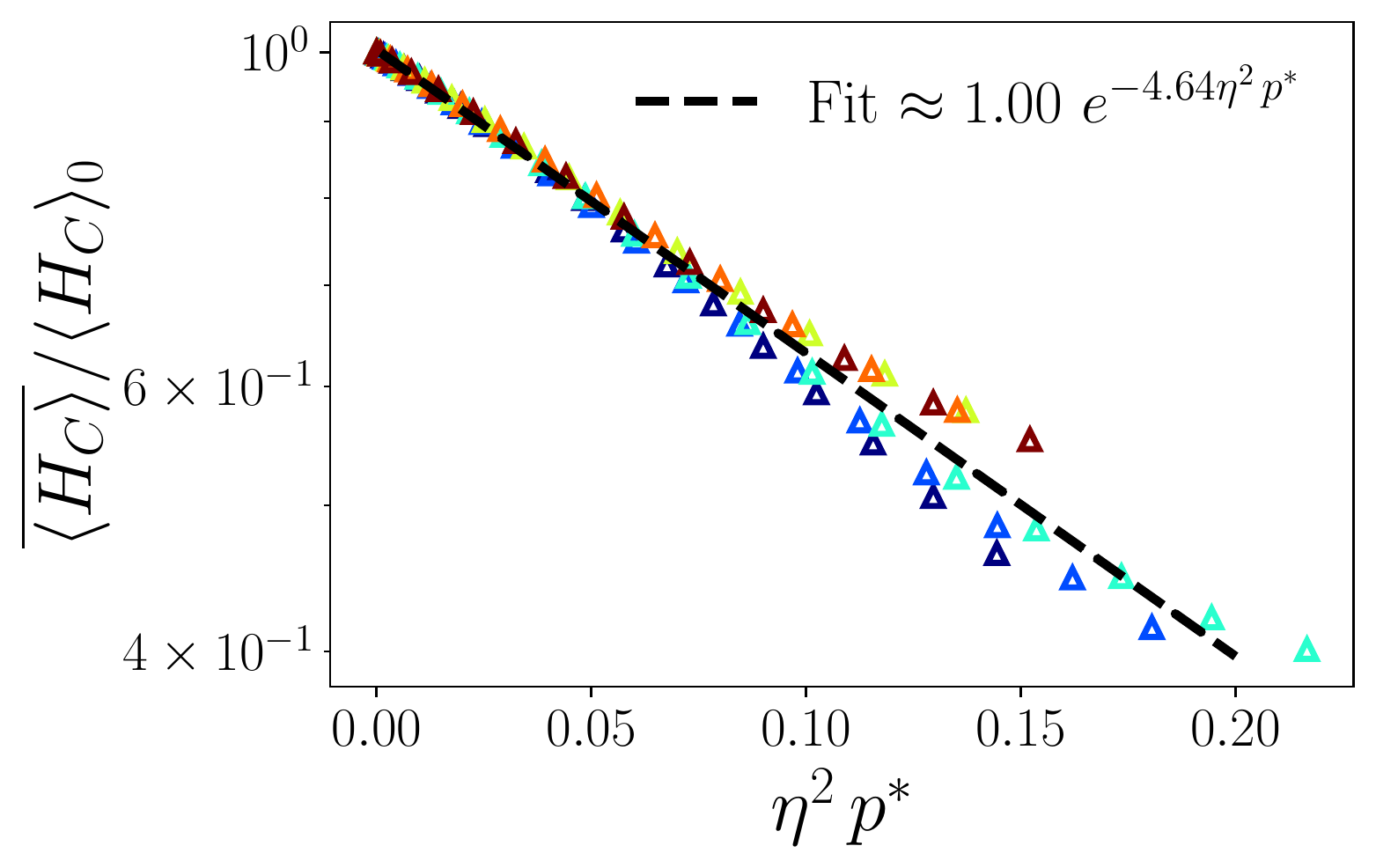}
    \caption{Scaling fits for the expectation value of the cost function for GHZ state preparation in the presence of coherent error. The decay is consistent with an exponential decay. }
    \label{fig:averagedOvercoherentErrorsIsing}
\end{figure}

\bibliographystyle{apsrev4-1}
\bibliography{refs}
\end{document}